\documentclass[11pt]{article}
\include{GrandMacros}

\usepackage[
top    = 2cm,
bottom = 0cm,
left   = 2.cm,
right  = 2.00cm]{geometry}

\usepackage{graphicx,verbatim,array,multicol, palatino}
\usepackage{amsthm, amsmath, amssymb,  setspace,natbib}
\usepackage{epsfig, url,epstopdf}
\usepackage{indentfirst,enumerate}
\usepackage{color}
\usepackage[super]{cite}
\usepackage[titletoc,title]{appendix}
\usepackage{booktabs}
\usepackage{graphicx}
\usepackage{subcaption}
\usepackage{adjustbox}
\usepackage{tabularx, booktabs}


\definecolor{red}{rgb}{1,0,0}
\definecolor{blue}{rgb}{0,0,1}
\definecolor{green}{rgb}{0,0.6,0.4}

\def \bbeta{\boldsymbol{\beta}}
\def \bLambda {\boldsymbol{\Lambda}}
\def \bSigma {\boldsymbol{\Sigma}}
\def \bU {\bf{U}}
\def \bu {\bf{u}}
\def \bZ {\bf{Z}}
\def \bI {\bf{I}}

\def \bY {\bf{Y}}
\def \bR {\bf{R}}

\def \bX {\bf{X}}

\title{A Geometric Perspective on the Power of Principal Component Association Tests in Multiple Phenotype Studies}
\author{Zhonghua Liu and Xihong Lin\thanks{Zhonghua Liu is Assistant Professor, Department of Statistics and Actuarial Science,  University of Hong Kong, Pokfulam, Hong Kong, China (Email: zhhliu@hku.hk).  Xihong Lin is Chair and Henry Pickering Walcott Professor of Biostatistics, Department of Biostatistics, Harvard T.H. Chan School of Public Health, Boston, MA 02115 (Email: xlin@hsph.harvard.edu).  Xihong Lin is also Professor of Statistics at Harvard University.  This work was supported by the National Institutes of Health grants CA R35-CA197449, P01-CA134294, U01-HG009088, U19-CA203654, and R01-HL113338. The authors thank the Associate Editor and  reviewers for their constructive comments, which helped improve the paper.}}
\date{}

\begin{document}
 \pagenumbering{gobble}
\maketitle

\begin{singlespace}
\noindent
\begin{abstract}
Joint analysis of multiple phenotypes can increase statistical power in genetic association studies. Principal component analysis, as a popular dimension reduction method,  especially when the number of phenotypes is high-dimensional, has been proposed to analyze multiple correlated phenotypes. It has been empirically observed that the first PC, which summarizes the largest amount of variance, can be less powerful  than higher order PCs and other commonly used methods in detecting genetic association signals. In this paper, we investigate the properties of PCA-based multiple phenotype analysis from a geometric perspective by introducing a novel concept called principal angle. A particular PC is powerful if its principal angle is $0^o$ and is powerless if its principal angle is $90^o$. Without  prior knowledge about the true principal angle, each PC can be powerless. We propose linear, non-linear and data-adaptive omnibus tests by combining PCs.  We demonstrate that the Wald test is a special  quadratic PC-based test. We show that the omnibus PC test is robust and powerful in a wide range of scenarios.  We study the properties of the proposed methods using power analysis and eigen-analysis. The subtle differences and close connections between these combined PC methods are illustrated graphically in terms of their rejection boundaries.  Our proposed tests have convex acceptance regions and hence are admissible. The $p$-values for the proposed tests can be efficiently calculated analytically and the proposed tests have been implemented in a publicly available R package {\it MPAT}.  We conduct simulation studies in both low and high dimensional settings with various signal vectors and correlation structures.  We apply the proposed tests to the joint analysis of metabolic syndrome related phenotypes with data sets collected from four international consortia to demonstrate the effectiveness of the proposed combined PC testing procedures. 
\end{abstract}

\vspace{1pc}
\end{singlespace}

\newpage
\pagenumbering{arabic}

\section{Introduction}
\label{intro}
In the past decade, Genome-Wide Association Studies (GWASs) have identified thousands of genetic variants associated with hundreds of human complex traits and diseases \citep{Welter2014}, as reported in  the National Human Genome Research Institute and European Bioinformatics Institute's (NHGRI-EBI) catalog. By using the open-access NHGRI-EBI catalog, \citet{Sivakumaran2011} found abundant evidence of pleiotropy: 233 (16.9\%) genes and 77 (4.6\%) single nucleotide polymorphisms (SNPs) show pleiotropic effects, and the numbers are still growing over time. As detailed phenotype data from epidemiological studies, electronic health records (EHR), genome-wide omics profiling and real-time mobile health devices are becoming rapidly available,  there is an increasing interest in  identifying cross-phenotype associations \citep{Solovieff2013,Bush2016}, which hold great potentials for  novel drug target discovery, drug repurposing and informing precision medicine \citep{Collins2015}.

 Our work is motivated by studying the genetic basis of metabolic syndrome (MetS) \citep{Brown2016}. A set of clinical phenotypes are involved in the disease process of MetS.  Single-trait GWAS studies have been conducted to identify susceptible SNPs associated with each of those MetS related phenotypes. The following four consortia studied the genetic architecture of the MetS traits. The International Consortium for Blood Pressure (ICBP) is an international effort to investigate blood-pressure genetics. It conducted a GWAS of Systolic Blood Pressures (SBP) of 200,000 individuals of European descent \citep{icbp2011genetic}. The Global Lipids Genetics Consortium (GLGC)  performed individual trait GWAS analysis of high-density lipoprotein cholesterol (HDL), low-density lipoprotein cholesterol (LDL) and triglycerides (TG) \citep{Teslovich2010}. It  examined the SNP-lipid associations  in 188,578 European-ancestry individuals \citep{Willer2013}. The Meta-Analyses of Glucose and Insulin-related traits Consortium (MAGIC) represents a collaborative effort to combine data from multiple GWASs to identify genetic loci that impact glycemic and metabolic traits.   The MAGIC study performed meta-analysis of  29 GWASs of Fasting Glucose (FG) from 58,074 non-diabetic participants,  and 26 GWASs of Fasting Insulin (FI)    from 51,750 non-diabetic participants \citep{manning2012}, with both analyses adjusting for Body Mass Index (BMI). The Genetic Investigation of ANthropometric Traits (GIANT) consortium investigates the genetic underpinning  that modulates human body size and shape.  It performed a GWAS analysis  of Body Mass Index (BMI)  using 339,224 individuals  \citep{locke2015genetic}, and a  BMI-adjusted GWAS analysis  waist-hip-ratio (WHR) using 224,459 individuals of European ancestry \citep{shungin2015}. Although those aforementioned studies  identified the SNPs associated with each of the eight  phenotypes,  the single-trait analysis paradigm  is likely to suffer from potential power loss for detecting the genetic variants associated MeS by ignoring the fact that these clinical phenotypes of MeS are related and might share a common genetic basis.

It has been shown that joint analysis of multiple phenotypes together can  increase statistical power to detect genetic variants. Numerous methods have been proposed for multiple phenotype analysis, see  \citet{Solovieff2013} for a review.  Examples 
include multivariate regression based methods, 
which  improve power under specific parametric assumptions, such as homogeneous effects across phenotypes, but are subject to power loss when these assumptions are violated \citep{Schifano2013, Zhou2014}; the $p$-value correction method TATES \citep{vanderSluis2013}, which accounts for between-phenotype correlation, has a good power in the presence of a very few association signals and can lose power otherwise. Furthermore, this method is subject to  inflated type I error rate by 12\% \citep{He2013}.  \citet{Zhu2015} proposed two  tests, one for detecting homogeneous effects and another for detecting heterogeneous effects based on a truncated test statistic. These tests were found to have good performance when their corresponding assumptions hold. In practice, researchers usually have little prior knowledge about which assumption holds, and hence it might be challenging to decide which test to use. Moreover,  the  $p$-value of the truncated test for detecting heterogeneous effects could not be calculated analytically and  requires Monte-Carlo simulations, which are computationally expensive for genome-wide analysis of multiple phenotypes.    \citet{Huang2013} and \citet{liu2017} proposed variance component tests for multiple phenotypes. We will show  that this variance component test is a special quadratic combination of  PC test in this paper. 

 Principal Component Analysis (PCA), as a popular dimension reduction technique,  especially when the number of phenotypes is not small, is an appealing approach that transforms correlated phenotypes into orthogonal composite scores \citep{Aschard2014}. Although it has been empirically found that principal components (PCs) that explain a small amount of the total variance of the multiple phenotypes can be as powerful or even more powerful than the PCs that explain a large amount of the total variance of the multiple phenotypes \citep{Aschard2014},  however there is no theoretical explanation for this  counter-intuitive phenomenon.  It is also unclear which PCs should be used to achieve the best power for genetic association testing. 

It is well known that the Uniformly Most Powerful (UMP) test does not exist for composite hypothesis testing. The classical Wald test can lose substantial power when the first PC captures all the signals and also explains a large amount of the total variance. The canonical correlation analysis aims to find a best linear combination of the multiple phenotypes \citep{Ferreira2009}, and thus can perform poorly when the relationship between a genetic variant and multiple phenotypes is not linear. Therefore, there is a  pressing need to develop effective powerful testing methods for multiple phenotype association studies.

Since multiple PCs are likely to contain association evidence, it could be  advantageous to combine PCs together to achieve better power. There are several challenges on how to effectively combine association evidence across multiple PCs.  First,
the underlying genetic effects are unknown and can be heterogeneous, i.e., a genetic variant can have positive, negative or null effects on different phenotypes. Second, the correlation structure among  multiple phenotypes can be arbitrary,  i.e.,  phenotypes can be positively or negatively correlated with varying correlation strength. However, little is known in the literature about the effect of the correlation structure on the power of the PC based tests.  Furthermore, it is more challenging to effectively combine PCs in high dimensional settings, such as in gene expression studies, because it is more complex to understand the interplay between the high dimensional signal vectors and the between-phenotype correlation structures. 
Therefore, it is of significant interest to develop more powerful  testing procedures by effectively combining PCs and taking into account the between-phenotype correlation structure, the effect size and the direction of the genetic effects, in both low and high dimensional settings.

In this paper, we aim to address these problems by developing robust and powerful PC-based methods for testing for genetic association with multiple phenotypes, as well as studying the  effects of the between-phenotype correlation structures on the power of the proposed PC-based tests. This paper makes the following contributions.   First, we introduce a novel geometric concept called principal angle and show that a particular PC can be powerless if its  principal angle is $90^o$ and can be as powerful as the Oracle test if its principal angle is zero. In practical settings, any PC can be powerless if one has no prior knowledge about the true principal angles.
	
Second,  we propose several data-driven  methods to combine PCs  to boost the power for testing for the association between a genetic variant and multiple phenotypes.  We first  propose the minimum  PC $p$-value (PCMinP) and the  Fisher's  method by combining PC $p$-values (PCFisher) as  testing statistics.   We then propose linear and quadratic combinations of PCs weighted by the functions of eigenvalues. Specifically, we show that an inverse-eigenvalue weighted linear combination of PCs (PCLC) can be as powerful as the Oracle test when all the principal angles are equal to each other, but can lose power otherwise.  Quadratic combinations of PCs are shown to be more robust than PCLC. We show that the classical Wald test and the recently proposed variance component score test \citep{Huang2013, liu2017} are  special cases of the quadratic combinations of PCs. These two tests both favor the alternatives under which the last principal angle is zero. As we usually have no prior knowledge about the true signal direction in practice, we propose an omnibus test  (PCO) which uses the data driven method to best combine several linear and   nonlinear  PC tests together to achieve robust power performance under various alternatives.  

Third, we perform eigen-analysis to investigate the effects of  the between-phenotype correlation structure on the power performance of the PC-based tests. The subtle differences and close connections between our proposed tests are compared graphically in terms of their rejection boundaries. Our proposed tests all have convex acceptance regions and hence are admissible \citep{Birnbaum1954, Birnbaum1955}. Fourth, the $p$-values of our proposed tests can be calculated analytically in a computationally efficient manner. 

The type I error rates of our proposed tests are shown to be well controlled by simulation studies. The powers of the proposed tests relative to several commonly used methods, such as the Wald test, the TATES method \citep{vanderSluis2013}, are compared using simulations in both low and high dimensional settings.  The robust power performance of the proposed omnibus test PCO is demonstrated through simulations using a range of signal patterns and correlation structures. 
Lastly, we applied our proposed tests to the aforementioned   metabolic syndrome trait GWAS data sets and identified  additional new genetic variants that were missed by the original univariate analyses. Those identified new SNPs might play important biological roles in the pathogenesis of MetS and can serve as potential candidates for future functional studies. 

The remainder of this paper is organized as follows. In Section \ref{sec_PCAT}, we describe our PC-based testing procedures and perform power analysis.  In Section \ref{sec_omni}, the omnibus PC-based tests are proposed to improve robustness and power of the PC-based tests.  In Section \ref{sec_rejectionBoundary}, we 
compare those tests in terms of their rejection boundaries and demonstrate their differences graphically.
In Section \ref{sec_Eigen}, we perform eigen-analysis to investigate how the between-phenotype correlation structure affects the statistical powers of our proposed tests.  In Section \ref{sec_sim}, we conduct  simulation studies to  evaluate the performance of our methods in both low and high-dimensional settings.  In Section \ref{sec_MetS}, we apply our tests to the  metabolic syndrome trait  GWAS data sets. Finally, we conclude with discussions in Section \ref{sec_discussion}.

\section[]{The Principal Component Association Tests}
\label{sec_PCAT}
Suppose that there are $K$ correlated phenotypes denoted by ${\bY}=(Y_1,\dots,Y_K)^T$.   Traditional GWAS studies consist of  hundreds of thousands of SNPs across the genome. One analyzes a SNP a time for each phenotype separately.  For a particular SNP, we have $K$ correlated test statistics for testing for the presence of genetic effects, i.e., Z-scores ${\bZ}=(Z_1,\dots,Z_K)^T$ that asymptotically follow a multivariate normal distribution with the covariance matrix $\bSigma$, which is equal to the
correlation matrix of $\bY$ conditional on other covariates included in the univariate analysis under the null \citep{liu2017}.  In other words, the $Z$ scores have already taken into account the effects of confounders, such as population stratification, and $\bSigma$ is not the crude covariance of $\bY$ but the residual covariance after regressing $\bY$ on covariates. Although across the whole genome, different genetic variants could have different minor allele frequencies (MAF), however their association test statistics $\bZ$ follow the same null distribution. This serves as the basis for consistently estimating $\bSigma$ using the sample covariance matrix of  the $\bZ$-statistics    across the genome under the null hypothesis \citep{Zhu2015, liu2017}.

For simplicity, we assume $\bSigma$  is known for the ease of discussions hereafter. For a given data set of sample size $n$, univariate analysis for each phenotype can be performed. For a particular genetic variant, we can obtain a $K$-dimensional  vector of summary testing statistics  ${\bZ} \sim N( \bbeta,{\bSigma})$, where $\bbeta \propto \sqrt{n}$ and $n$ is the sample size for calculating $\bZ$.  We are interested in testing $H_0$: $\bbeta = 0$  against $H_a$: $ \bbeta\neq 0$, where $\bbeta$ is referred to as the signal vector. We would like to develop robust and powerful  tests  that are robust to the between-phenotype correlation structures and signal vector patterns, especially when the dimension of phenotypes is not small.

\subsection{The Oracle Test for the Fixed Alternative Hypothesis}
Under the fixed alternative hypothesis $\bbeta$,  the Uniformly Most Powerful (UMP) test is
\begin{equation}\label{Oracle}
\mbox{Oracle} = {\bbeta}^T{\bSigma}^{-1}{\bZ},
\end{equation}
which directly follows from the Neyman-Pearson Lemma \citep{Bittman2009}. One can easily see that the Oracle test is a linear combination of $\bZ$ with the coefficients  depending on the true  $\bbeta$ and $\bSigma$. It is natural to view this hypothesis testing problem as a binary classification problem. We observe a vector $\bZ$ and need to decide whether $\bZ$ is from the null $H_0$ or the alternative $H_a$. This classification problem fits into the framework of linear discriminant analysis (LDA). In fact, this Oracle test can be viewed as the Fisher LDA \citep{Fisher1936},  which is the Bayes optimal classifier \citep{Bickel2004} and provides the highest sensitivity uniformly at any given specificity \citep{Su1993}. In practice, we do not know the true ${\bbeta}$ and therefore we cannot perform this Oracle test. Nonetheless, we can  use it as an { \it ideal} benchmark for power comparisons with those implementable tests. 

 Under the alternative hypothesis $H_a$: $\bbeta\neq 0$,    both $\bbeta$  and correlation matrix $\bSigma$ are unknown.  Equation (\ref{Oracle})  implies that only a `smart' linear combination of individual $Z$-testing statistics   that is as close as possible to the unknown true quantity $\bSigma^{-1}\bbeta$, can be as powerful as the Oracle test,  but at the potential risk of being powerless if the linear combination is not `smart'.

\subsection{Single Principal Component Tests for the Composite Hypothesis}
\label{sec_singlePC}
 Consider the composite hypothesis $H_0$: ${\bbeta} =0$ versus $H_a$: ${\bbeta}\neq 0$. 
Using spectral decomposition, we have
\[
{\bSigma} = {\bU}{\bLambda} {\bU}^T = \sum_{k=1}^K \lambda_k{{\bu}_k{\bu}_k^T},
\]
where $\bLambda$ is a diagonal matrix whose diagonal elements are the eigenvalues $\lambda_1\geq \lambda_2 \geq \dots \geq \lambda_K > 0$ of $\bSigma$, and $\bU$ is the normalized orthogonal matrix whose $k$th column ${\bf{u}}_k$ is the $k$th eigenvector associated with the $k$th largest eigenvalue $\lambda_k$ of $\bSigma$. We also require that $\bU$ is a proper rotation matrix, that is $\det(\bU)=1$ \citep{Pettofrezzo1978}. The $K$ eigenvectors ${\bu}_k$ $(k=1,\dots,K)$ form an eigen-basis and hence constitute a new orthogonal coordinate system in which the $k$th  coordinate direction  corresponds to the $k$th principal component $\mbox{PC}_k$. It is straightforward to show that the  distribution of $\mbox{PC}_k$ is
\[
\mbox{PC}_k = {\bu}^T_k {\bZ} \sim N({{\bu}_k^T \bbeta},\lambda_k) , 1\leq k \leq K.
\]

As $||{\bu}_k||^2=1$,  the non-centrality parameter ($ncp$) of $\mbox{PC}_k$ under the alternative  is
\begin{equation*}
ncp_k = \frac{({\bu}_k^T{\bbeta})^2}{\lambda_k}=\frac{||{\bbeta}||^2\{\cos(\theta_k)\}^2}{\lambda_k},
\label{ncp_PC}
\end{equation*}
where $\theta_k  \in [0,180^o]$ is the angle  between the signal vector ${\bbeta}$ and the eigenvector ${\bu}_k$  and is defined as the $k$th Principal Angle (PA), and $||{\bbeta}|| = \sqrt{\sum_{k=1}^K\beta_k^2}$ which is defined as the overall signal magnitude.  An underlying  constraint for the principal angles is that $\sum_{k=1}^K\cos^2(\theta_k)=1$, which will be used for power analysis later. 
If the $k$th principal component $\mbox{PC}_k$ is used as a testing statistic for
$H_0$: ${\bbeta} =0$ versus $H_a$: ${\bbeta}\neq 0$, then its theoretical power at significance level $\alpha$ is
\begin{equation*}
\mbox{Power}=\Phi(Z_{\frac{\alpha}{2}} +\sqrt{ncp_k}) +\Phi(Z_{\frac{\alpha}{2}} -\sqrt{ncp_k}),
\end{equation*}
where $\Phi(\cdot)$ is the cumulative standard normal distribution function, and $Z_{\frac{\alpha}{2}}$ is its  $\alpha/2$ percentile.
If $\theta_k=0$, then $\mbox{PC}_k$ is as powerful as the Oracle test; however if $\theta_k=90^o$, then $\mbox{PC}_k$ is powerless. This geometric perspective clearly explains why using any particular PC could be powerless to detect association  signals in the situations where its principal angle is $90^o$.

The Principal Angle of a PC measures the the degree of similarity between the direction of the PC of multiple phenotypes and the direction of the genetic effect vector of a SNP on multiple phenotypes.  When the principal angle of a PC is zero, it means that the direction of the PC completely aligns with the genetic effect direction, and is thus perfect for being used for summarizing multiple phenotypes into a scalar super-phenotype for detecting genotype-phenotype associations. If the principal angle of a PC is 90 degree, it means that the PC contains no information about the genetic effects and is thus not useful for detecting the genotype-multiple phenotype associations.

The power analysis for single PC test serves as the building blocks of the power analysis of combined PC based tests. We  observe that the power of single PC test depends not only on $\bbeta$ but also  $\bSigma$ through its eigenvalues and eigenvectors,  which will be investigated by eigen-analysis in Section \ref{sec_Eigen}.  It should be noted that the PC directions of the Z-scores are often not the same as the PC directions of the original phenotypes $\bY$, as the Z-scores have taken the confounders into account.


\subsection{The PCMinP Test}
As the signal vector $\bbeta$ is unknown in practice,  one usually has no prior information about the true principal angles and thus cannot decide which PC to use for association testing. Hence,  we propose to use the minimum principal component $p$-value  as a testing statistic named PCMinP, 
\[
\mbox{PCMinP} = \min_{1\leq k \leq K} p_k,
\]
where $p_k$ is the $p$-value based on $\mbox{PC}_k$. In fact, PCMinP is equivalent to using $\sup _{1\leq k \leq K}|\mbox{PC}_k|/\sqrt{\lambda_k}$ as a test statistic, and hence can be viewed as a nonlinear combination of PCs.  Because the $K$ PCs are mutually independent, so the $p$-value of PCMinP  can be easily computed  as $p = 1-(1-\mbox{PCMinP})^K$.

Denote $\alpha^* = 1-(1-\alpha)^{1/K}$ where $\alpha$ is a pre-specified significance level, then the power of PCMinP under the alternative is
\begin{eqnarray*}
\mbox{Power}=1-\prod_{k=1}^K \left[ 1-\left\{ \Phi(Z_{\frac{\alpha^*}{2}} +\sqrt{ncp_k}) +\Phi(Z_{\frac{\alpha^*}{2}} -\sqrt{ncp_k}) \right \}\right ].
\end{eqnarray*}
Suppose that $||{\bbeta}||$ and $\lambda_k$ are fixed, then the power of PCMinP is maximized when $\theta_K=0$ and its maximal power is
\begin{equation*}
\mbox{Power}_{\max} = 1-(1-\alpha)^{\frac{K-1}{K}}\left\{ 1-  \Phi\left(Z_{\frac{\alpha^*}{2}} +\frac{||\boldsymbol \beta||}{\sqrt{\lambda_K}}\right) -\Phi\left(Z_{\frac{\alpha^*}{2}} -\frac{||\boldsymbol \beta||}{\sqrt{\lambda_K}}\right) \right\}.
\end{equation*}
This implies that PCMinP favors the alternatives under which the last PC captures all the signals. Furthermore, the power of PCMinP goes to 1 as $\lambda_K \rightarrow 0$. The power of PCMinP is minimized when  $\cos^2(\theta_k)=\lambda_k/K$ ($k=1,\dots, K$), and the minimum power is
\begin{equation*}
\mbox{Power}_{\min} = 1-\left\{ 1-  \Phi\left(Z_{\frac{\alpha^*}{2}} +\frac{|| \bbeta||}{\sqrt{K}}\right) -\Phi\left(Z_{\frac{\alpha^*}{2}} -\frac{|| \bbeta||}{\sqrt{K}}\right) \right\}^K.
\end{equation*}
The result follows directly from the inequality of arithmetic and geometric means. This implies that the worst situation for PCMinP is that  all the PCs are equally powerful,  e.g., when multiple phenotypes are independent, e.g., $\bSigma=\bI$. 

\subsection{The PCFisher Test}
PCMinP aims to pick the most powerful PC direction and discards the other less powerful PCs. Hence PCMinP does not fully use  available information contained in all the PCs.  We hereby propose to combine all the $K$ independent principal component $p$-values using Fisher's method~\citep{Fisher1932} with its null distribution  given by
\[
\mbox{PCFisher} = -2\sum_{k=1}^K\log(p_k) \sim \chi^2_{2K}.
\]
PCFisher can also be viewed as a nonlinear combination of the PCs,
\begin{equation}\label{PCFisher_fun}
\mbox{PCFisher} = -2\sum_{k=1}^K\log\{1-F_{\chi_1^2}(\mbox{PC}_k^2/\lambda_k)\},
\end{equation}
where $F_{\chi_1^2}(\cdot)$ represents the chi-squared cumulative distribution function with one degree of freedom. Equation (\ref{PCFisher_fun}) implies that PCFisher allocates larger weights to PCs with smaller eigenvalues. Therefore,
PCFisher achieves its maximal power when $\theta_K=0$ and achieves its minimal power when $\theta_1=0$ for fixed $|| \bbeta||$ and $\lambda_k$. The
Fisher's $p$-value combination method is asymptotically Bahadur optimal (ABO), in the sense that the $p$-value of PCFisher converges to zero  with the fastest rate under the alternative when the sample size goes to infinity \citep{Bahadur1967,Littell1971, Littell1973}.

\subsection{The Test Based on a Linear Combination of PCs}
\label{sec_PCLC}
Motivated by the inverse variance weighting method commonly used when combining independent tests \citep{mosteller1954,liptak1958}, we obtain the following linear combination of PCs with each PC weighted by its inverse variance, 
\[
\mbox{PCLC} = \sum_{k=1}^K \frac{\mbox{PC}_k}{\lambda_k} \sim N(0,\sum_{k=1}^K\lambda^{-1}_k).
\]
Under the alternative hypothesis,  its non-centrality parameter is
\[
ncp = \frac{\left( \sum_{k=1}^K\lambda_k^{-1}{\bf{u}^T_k{ \bbeta}}\right)^2}{\sum_{k=1}^K\lambda_k^{-1}}=\frac{||{ \bbeta}||^2\left \{ \sum_{k=1}^K\lambda_k^{-1} \cos(\theta_k)\right \}^2}{\sum_{k=1}^K\lambda_k^{-1}}.
\]
We now study when the power of PCLC will be maximized and minimized with respect to $\theta_k$,  for any fixed $||{ \bbeta}||$ and $\lambda_k$. This can be formulated as the following constrained optimization problem
\begin{equation*}
\begin{array}{rrclcl}
\displaystyle \max_{\cos(\theta_k)} & \multicolumn{3}{l}{\sum_{k=1}^K\lambda_k^{-1} \cos(\theta_k)} \\
\textrm{s.t.} & \sum_{k=1}^K\cos^2(\theta_k) & = & 1. \\
\end{array}
\end{equation*}
Using the Lagrange multiplier method, we obtain that the power of PCLC is maximized when the principal angles satisfy the following conditions
\begin{equation}
\cos^2(\theta_k) = \frac{\lambda_k^{-2}}{\sum_{k=1}^K\lambda_k^{-2}}, k=1,\dots, K.
\label{PCLC_max}
\end{equation}
In fact,  we can rewrite PCLC as $\mbox{PCLC} = (\bU\bLambda^{-1}{\bf J})^T{\bf Z}$,  where ${\bf J} = (1,\dots,1)^T$. Hence, PCLC  will achieve its own maximal power when ${\bbeta} \propto {\bU}{\bLambda}^{-1} {\bf J}$, which is equivalent to equation (\ref{PCLC_max}). It can be easily seen that PCLC 
is powerless when ${\bbeta} \perp {\bU}{\bLambda}^{-1} {\bf J}$ where positive and negative genetic effects are canceled out. The PCLC test can be as powerful as the Oracle test when ${\bbeta} \propto \bU {\bf J}$, or equivalently when all the principal angles are equal to each other, i.e., $\cos^2(\theta_k) = 1/K (k=1,\dots,K)$. In other words, when all the $K$ principal angles are the same, PCLC is more powerful than any other tests to detect such alternatives.  

\subsection{Quadratic Combination of PCs}
PCLC is very sensitive to principal angles and can be powerless as shown theoretically in Section \ref{sec_PCLC} and empirically in the simulation setting M5 in Table \ref{power_result}. To overcome this drawback, we propose to combine PCs using the following weighted quadratic function  that weights the PCs by a function of the eigenvalues.
\begin{equation*}
\mbox{PCQ}_{\gamma} = \sum_{k=1}^K \lambda_k^{1-\gamma}\left(\frac{\mbox{PC}_k}{\sqrt{\lambda_k}}\right)^2,
~  0 \leq \gamma < +\infty,
\label{weighted_Quad_PC}
\end{equation*}
where $\gamma$ controls the relative importance of each PC in the quadratic combinations. For example, if $\mbox{PC}_1$ captures most of the signals, then we can choose smaller $\gamma$; while if $\mbox{PC}_K$ captures most of the signals, then we can choose larger $\gamma$.  Let ${\bf K}_{\gamma} = {\bU} {\bLambda}^{-\gamma}{\bU}^T $ and denote the transformation as
$\phi_{\gamma}({\bf Z}) = {\bU}^T{\bZ}/\lambda^{\gamma/2}$, then $\mbox{PCQ}_{\gamma}$ can be rewritten as $\mbox{PCQ}_{\gamma} = {\bZ}^T {\bf K}_{\gamma} {\bZ} = \langle\phi_{\gamma}({\bZ}),\phi_{\gamma}({\bf Z})\rangle$, where $\langle\cdot,\cdot\rangle$ denotes the inner product in the transformed feature space.
From this point of view, $\mbox{PCQ}_{\gamma}$ is a kernel based testing statistic 
\citep{liu2007semiparametric}. The choice of $\gamma$ is essentially a choice of kernel and reflects our prior belief in the true alternative. We show in this section that several commonly used tests with $\gamma=0,1,2$ are special cases of $\mbox{PCQ}_{\gamma}$.

When $\gamma = 0$, $\mbox{PCQ}_{\gamma}$ has the following form
\[
\mbox{WI} = \mbox{PCQ}_0= \sum_{k=1}^K \mbox{PC}_k^2 = \sum_{k=1}^K Z_k^2,
\]
which follows from the fact that $\bU$ is an isometric transformation and ${\bU} {\bU}^T = \bI$. This choice of $\gamma$ assumes a working independence (WI) relationship among the $K$ Z-scores since ${\bf K}_{\gamma}$ reduces to an identity matrix.  Under the null, WI  follows a  mixture of chi-squared distribution $\sum_{j}\lambda_j\chi_{1j}^2$, where $\lambda_j$ are the eigenvalues of $\bSigma$ and $\chi_{1j}^2$ are independent $\chi_1^2$ random variables. Hence, its $p$-value can be computed using the exact method~\citep{Davies1980a}.  At the significance level $\alpha$, we reject the null hypothesis $H_0$: $ \bbeta = 0$ if $\sum_{k=1}^K \mbox{PC}_k^2 > C_{\alpha}$ where $P(\sum_{k=1}^K \mbox{PC}_k^2 > C_{\alpha};H_0) =\alpha$. Thus, the acceptance region of WI  is a $K$-dimensional ball with radius equal to $\sqrt{C_{\alpha}}$. Although the acceptance region of WI is spherically symmetric, however the probability distribution of $\bZ$ is not spherically symmetric unless $\bSigma$ is an identity matrix. Under the alternative, the power of WI favors the alternatives under which $\mbox{PC}_1$ captures all the signals. This is because $\mbox{PC}_1$ has the largest variance and hence signals from the $\mbox{PC}_1$ direction are more likely to fall outside of this ball-shape acceptance region  (See Figure \ref{PCA_rejectionBoundary}).

When $\gamma = 1$, $\mbox{PCQ}_{\gamma}$ becomes the classical Wald test as
\[
\mbox{Wald}= \mbox{PCQ}_1=\sum_{k=1}^K\frac{\mbox{PC}_k^2}{\lambda_k} \sim \chi^2_K.
\]
This can be easily shown using the fact that  ${{\bf Z}^T \bf{\Sigma^{-1}}\bf{Z}} = {( {\bU}^T {\bZ})^T{\bLambda}^{-1}({\bU}^T{\bZ})}= \sum_{k=1}^K \mbox{PC}_k^2/\lambda_k.$ At the significance level $\alpha$, its acceptance region is determined by
 $\sum_{k=1}^K\mbox{PC}_k^2/\lambda_k \leq C_{\alpha}$, which is a $K$-dimensional ellipsoid and $C_{\alpha}$ is the $1-\alpha$ percentile of $\chi^2_K$. Under the alternative, the distribution of the Wald test is a non-central chi-squared distribution with non-centrality parameter $ncp =\sum_{k=1}^K ||{\bbeta}||^2 \cos^2(\theta_k)/\lambda_k $. To know when the Wald test achieves its maximal power for any fixed $||{\bbeta}||$ and $\lambda_k$, we can solve the following constrained optimization problem
\begin{equation*} \label{PCSS_power}
\begin{array}{rrclcl}
\displaystyle \max_{\cos^2(\theta_k)} & \multicolumn{3}{l}{\sum_{k=1}^K\lambda_k^{-1} \cos^2(\theta_k)} \\
\textrm{s.t.} & \sum_{k=1}^K\cos^2(\theta_k) & = & 1. \\
\end{array}
\end{equation*}
Using standard linear programming technique, one can easily show that the power of the Wald test  is maximized when $\theta_K=0$,  i.e, when signals lie in the last PC direction, and minimized when $\theta_1=0$,  i.e., when signals lie in the first PC direction. Again, even though the Wald test achieves its maximal power when the last PC captures all the signals, this does not imply the Wald test is more powerful than its competitors   under such alternatives.

When $\gamma = 2$,  $\mbox{PCQ}_{\gamma}$ is
\[
\mbox{VC} = \mbox{PCQ}_2= \sum_{k=1}^K \frac{\mbox{PC}_k^2}{\lambda^2_k}.
\]
In this case, $\mbox{PCQ}_{\gamma}$ corresponds to the variance component (VC) score test $\mbox{VC} = {\bZ}^T{\bSigma}^{-1}{\bSigma}^{-1}{\bZ}$ \citep{Huang2013, liu2017}, which assumes that the $\beta_k$ ($k=1,\cdots, K)$ follow a common distribution with mean 0 and variance $\tau$ and tests for $H_0$: $\tau=0$. The equivalence between VC and $\mbox{PCQ}_2$ can be seen by observing that
$\mbox{VC} = {\bZ}^T {\bU}{\bLambda}^{-2}{\bU}^T {\bZ} = \sum_{k=1}^K \mbox{PC}_k^2/\lambda^2_k$. Compared with the Wald test,  VC gives even more weight to the last PC and hence is more powerful than the Wald test when $\theta_K=0$.  VC follows a mixture of chi-squared distributions $\sum_{k=1}^K {\lambda^{-1}_k}\chi^2_{1k}$ under the null, where $\lambda_k$ are the eigenvalues of $\bSigma$ and $\chi_{1k}^2$ are independent $\chi_1^2$ random variables, so its  $p$-value can be  computed using the exact method~\citep{Davies1980a}.
The acceptance region of VC is also a $K$-dimensional ellipsoid but has a different shape from that of Wald as shown in Figure \ref{PCA_rejectionBoundary} in Section \ref{sec_rejectionBoundary}.

Here, we present a simple example to illustrate the power difference between the three quadratic tests: WI, Wald and VC. Suppose we have a bivariate normal Z-scores  with correlation $\rho=0.8$.  The first eigenvector is ${\bf u}_1 = (1/\sqrt{2},1/\sqrt{2})$ and the second eigenvector is ${\bf u}_2 = (-1/\sqrt{2},1/\sqrt{2})$ by direct calculation. If ${\bbeta}=(2.5,2.5)^T$ which is in the direction of ${\bu}_1$, then the powers of WI, Wald and VC are $0.75,0.65,0.09$ respectively; and if ${\bbeta}=(-0.8,0.8)^T$ which is in the direction of ${\bu}_2$, then the powers of WI, Wald and VC are $0.09,0.61, 0.71$ respectively. This shows that the Wald test is less powerful than WI when the first PC captures all the signals, and the Wald test is less powerful than VC when the last PC captures all the signals. More power comparisons among those three tests are provided in Table \ref{power_result}. 

\section{The Omnibus PC-Based Tests}
\label{sec_omni}
\subsection{Adaptive Quadratic Combination of PCs}
The results in Section 2.6 show that a lack of prior knowledge about the true principal angle can lead to an unwise choice of $\gamma$, and the resulting test might have little power to detect the alternative.  Therefore, we propose to choose $\gamma$ in a data-adaptive fashion  by choosing  $\gamma$ using the data that yields the smallest $p$-value, and then use this smallest $p$-value as a test statistic.  In practice, it is computationally expensive to perform an exhaustive search for the optimal $\gamma$ in the whole range.
Instead, we restrict our search within $\gamma \in \{0,1,2\}$ and then pick the smallest $p$-value among WI, Wald and VC as a testing statistic named PCAQ
\begin{equation*}
\mbox{PCAQ} = \min_{\gamma} p_{\gamma},
\end{equation*}
where $p_{\gamma}$ is the $p$-value of $\mbox{PCQ}_{\gamma}$ for a given $\gamma$.  Note that WI, Wald and VC tests are correlated as they are calculated using the same data. Hence their $p$-values $p_{\gamma}$ are correlated. Calculations of the $p$-value of PCAQ  need to take their correlations into account. Specifically, the $p$-value of PCAQ can be calculated as
\begin{equation}\label{PCAQ_pval}
p = 1 - P\{\min_{\gamma}X_{\gamma} > \Phi^{-1}(\mbox{PCAQ})\},
\end{equation}
where $X_{\gamma} = \Phi^{-1}(p_{\gamma})$ and $\Phi^{-1}(\cdot)$ denotes the inverse standard normal cumulative distribution function.  

Equation (\ref{PCAQ_pval}) can be efficiently computed using the following multivariate normal distribution function  that has been implemented in the FORTRAN language \citep{Genz1992, Genz1993} and also wrapped  in the R package {\it mvtnorm} \citep{Genz2009}. This computation requires an input of the correlation matrix $\bR_{X}$ of the vector $(X_{\gamma=0},X_{\gamma=1},X_{\gamma=2})$ which only needs to be estimated once for the whole genome 
by the following algorithm:
\begin{enumerate}
\item Generate $B$ random samples from $\bf{Z}\sim N(\bf{0},\Sigma)$.
\item Compute the $p$-values of PCQ on the $b$th sample for $\gamma=0,1,2$, $~1\leq b \leq B.$
\item Perform inverse-normal transformation $X_{\gamma}=\Phi^{-1}(p_{\gamma})$ on the $b$th sample, where $\gamma=0,1,2$.
\item Take the sample correlation matrix $\widehat{\bR}_{X}$ across the $B$ realizations of $\bX_{\gamma}$.
\end{enumerate}
In practice, one can take $B=1000$ and this algorithm can provide a good estimate of ${\bR}_X$ (in a few seconds) which can be used for computing the $p$-values for millions of SNPs in the whole genome.

\subsection{The Omnibus PC-based Test}
The PCAQ test aims at constructing an optimal quadratic PC-based test. To construct an omnibus test across linear, quadratic and other non-linear tests, we can combine all the PC combination methods including PCMinP, PCFisher, PCLC, WI, Wald and VC together by taking the minimum $p$-value of them as the omnibus test statistic named PCO. The $p$-values  of those six tests are correlated as they are calculated using the same data. Similar to PCAQ, the $p$-value of PCO can also be computed by first performing an inverse-normal transformation of the $p$-value of the test statistic under consideration, then  using a multivariate normal distribution function  with the correlation matrix estimated using the same fast Monte Carlo simulation method described above. Compared to PCAQ, PCO combines three more non-quadratic tests and is expected to be more robust  than PCAQ for various alternatives. However, a price PCO has to pay for combing more tests is that it  might be slightly less powerful than PCAQ  when quadratic combinations of PCs already have good power, for example in the simulation setting M3 in Table \ref{power_result}. PCO is expected to be  more powerful than PCAQ when any of PCLC, PCFisher or PCMinP has better power than the quadratic combinations of PCs to detect the signals, as demonstrated in the simulation settings M4, M7, M12, M13 and M15 in Table \ref{power_result}.

\section{Comparison of the Rejection Boundaries of the PC-Based Tests}
\label{sec_rejectionBoundary}
\begin{figure}
\centerline{\includegraphics[scale=0.42]{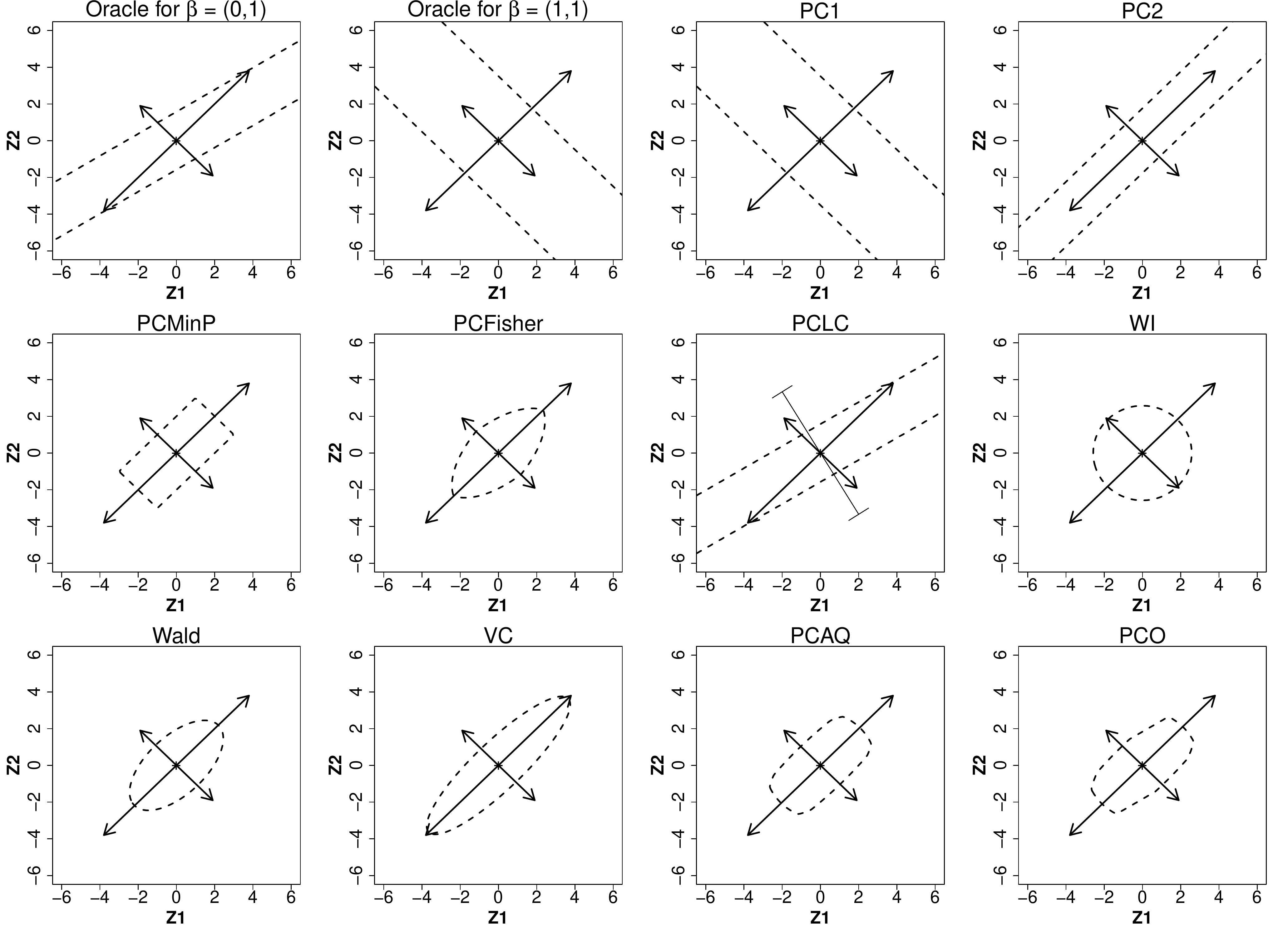}}
\caption[PCA_Rejection]{The rejection boundaries of the proposed PC-based tests for bivariate normal test statistics $(Z_1,Z_2)$ with correlation equal to $0.6$. The  dashed lines or curves represent the boundaries that separate the acceptance and rejection regions at the significance level $0.05$. The longer solid lines with  arrows represent the $\mbox{PC}_1$ direction, the shorter  solid lines with arrows represent the $\mbox{PC}_2$ direction, and the lengths of the longer and shorter solid lines with arrows are equal to $6\sqrt{\lambda_1}$ and $6\sqrt{\lambda_2}$ respectively. For PCLC,  the added solid line with ``T"-type arrows illustrates the direction for alternative ${\bbeta}$ which is orthogonal to its rejection boundaries, where $\theta_1=76^o$ and $\theta_2=14^o$. }
\label{PCA_rejectionBoundary}
\end{figure}

 In this section, we  compare the proposed PC-based tests graphically in terms of their  rejection boundaries. For the ease of  illustration, we focus on the  two dimensional $(Z_1,Z_2)^T$ space as given in Figure \ref{PCA_rejectionBoundary}. We also included the Oracle test for  ${\boldsymbol \beta}=(1,1)^T$ and ${\bbeta}=(0,1)^T$  assuming the true alternative is known. We set the correlation to be $0.6$, so  the two eigenvalues are $\lambda_1 = 1.6$ and $\lambda_2 = 0.4$, and the two corresponding eigenvectors are ${\bf u}_1 = (\frac{1}{\sqrt{2}},\frac{1}{\sqrt{2}})^T$ and ${\bf u}_2 = (-\frac{1}{\sqrt{2}},\frac{1}{\sqrt{2}})^T$ respectively. Suppose that the true alternative is ${\bbeta}=(1,1)^T$ which is in the same direction of ${\bu}_1$, then $PC_1$ has the same rejection boundary as the Oracle test. Suppose that the true alternative is ${\bbeta}=(0,1)^T$ which is in the same direction of ${\bU}{\bf J}$ where ${\bU}=({\bf u}_1,{\bf u}_2)$ is the eigenvector matrix  and ${\bf J} = (1,1)^T$, then PCLC has the same rejection boundary as the Oracle test. One can further deduce that $\mbox{PC}_2$ will have the same rejection boundary as the Oracle test if the true alternative is proportional to ${\bu}_2$. We  observe that all the proposed tests have convex acceptance regions. Hence the proposed tests are all admissible \citep{Birnbaum1954,Birnbaum1955}. This implies that each test can be more powerful than its competitors for  some alternatives but less powerful for others.

The rejection boundaries of $\mbox{PC}_1$ and $\mbox{PC}_2$ are all straight lines but are orthogonal to each other, indicating that these two PCs aim to detect orthogonal alternatives. $\mbox{PC}_2$ has a narrower gap between the two rejection boundary lines than that of $\mbox{PC}_1$, because $\mbox{PC}_2$ has a smaller variance (eigenvalue).
The rejection boundaries of PCLC are also straight lines but are not orthogonal to either $\mbox{PC}_1$ or $\mbox{PC}_2$. The angle between the rejection boundary lines  of PCLC  and $\mbox{PC}_1$ is $14^o$. Hence, if the mean vector ${\bbeta}$ also has angle $14^o$ with $\mbox{PC}_1$  direction, then  ${\bbeta}$ is parallel to the rejection boundaries of PCLC and will never be detected by PCLC. If ${\bbeta}$ has angle $76^o$ with $\mbox{PC}_1$  direction as shown by the solid line with ``T"-type arrows, then ${\bbeta}$ is orthogonal to the rejection boundaries of PCLC (shortest distance to the null) and will be detected by PCLC with its maximal power. However, this does not imply PCLC is more powerful than its competitors to detect the alternatives in the direction of the solid line with ``T"-type arrows because PCLC is not as powerful as the Oracle test for such alternatives.

The rejection boundary of PCMinP is a tilted rectangle with the edge lengths proportional to $\sqrt{\lambda_k},k=1,2$. PCMinP achieves its maximal power when ${\bbeta}$ is in the $\mbox{PC}_2$ direction (shortest distance to the null), while achieves its minimal power when  ${\bbeta}$ points to the four corners, under which $\mbox{PC}_1$ and $\mbox{PC}_2$ have equal powers. The rejection boundaries of Wald and VC are both ellipses. However, the minor axis of VC is shorter than that of Wald, while the major axis of VC is longer than that of Wald. This implies that  VC is more powerful than Wald when ${\bbeta}$ is in the $\mbox{PC}_2$ direction. The rejection boundaries of PCFisher are similar to that of Wald, which well explains why they have similar powers as will be demonstrated in the simulation studies. The rejection boundary of WI is a circle  and WI actually favors alternatives in the  $\mbox{PC}_1$ direction because it's more likely for the signals to fall outside of the rejection boundary of WI along the $\mbox{PC}_1$ direction compared to other PC directions. 
 
We observe that the rejection boundaries of PCMinP, PCAQ and PCO resemble each other because these three tests all use the minimum $p$-value as testing statistics across certain sets of tests and hence are data adaptive.  The rejection boundary of PCAQ is smooth and does not have sharp angles like that of PCMinP. The rejection boundary of PCO is more bumpy than that of PCAQ since it combines linear and non-linear tests.  The rejection boundary comparisons well explain the differences and connections between the proposed PC-based tests,  and illustrate that there is no UMP test for all the $\bbeta$ directions.

\section{Eigen-Analysis of Correlation Matrices and Their Effects on the PC-Based Tests}
\label{sec_Eigen}
The results in Sections \ref{sec_PCAT} and \ref{sec_omni} show that the powers of the PC-based tests depend on the principal angles $\theta_k$, the   eigenvalues $\lambda_k$ for a fixed norm of $\bbeta$. To test for associations between a SNP  and a set of multiple phenotypes, a question of practical interest is that  how the PC-based tests perform in the presence of a mixture of signal and noise phenotypes, especially when signals are sparse. For example, in studying the  effects of a SNP on a genetic pathway/network consisting of multiple gene expressions, it is common that a SNP affects some gene expressions but not others in the genomic pathway/network. In this section, we investigate how the correlation structure of $\bSigma$ affects its eigenvalues and eigenvectors, and subsequently affects the powers of PC-based tests. Suppose that $K_1$ out of $K$ phenotypes are associated with a genetic variant and $K_0=K-K_1$ of them are not, i.e., the  signal vector ${\bbeta}$ contains $K_1$ non-zeros (signals) and $K_0$ zeros (noises), denoted as
${\bbeta}^T =(\beta_1,\dots,\beta_{K_1},0,\dots,0)$.  For $\bSigma$, we consider the following partitioned correlation matrix 
\begin{equation*}
{\bSigma} =\left(\begin{array}{cc}%
{ \bSigma}_1 & {\bSigma}_2\\
{\bSigma}^T_2 &{\bSigma}_3
\end{array}\right),
\label{block_matrix}
\end{equation*}
where ${\bSigma}_{1}$ and ${\bSigma}_{3}$ denote the correlation matrices among signal and noise phenotypes respectively, and the $(i,j)$th element of ${\bSigma}_{2}$ denotes the correlation between the $i$th signal phenotype and the $j$th noise phenotype, $1\leq i \leq K_1$, $1\leq j \leq K_0$. We first obtain eigen-analysis results for special structured correlation matrices $\bSigma$, and then consider more general situations.

\vspace{-0.1in}
\subsection{Exchangeable Correlation Matrices}
\label{sec_eigen_exch}
If $\bSigma$ is exchangeable with correlation $\rho >0$,   then its eigenvalues are
\[
\lambda_1 = (K-1)\rho + 1; ~\lambda_k = 1 -\rho,~k=2,\dots, K,
\]
where the algebraic multiplicities of $\lambda_1$  and $\lambda_k$ are one and  $K-1$ respectively. This implies that the eigenspace associated with $\lambda_1$ is of dimension one and can be spanned by eigenvector ${\bf{u}}^T_1 = (\frac{1}{\sqrt{K}},\frac{1}{\sqrt{K}}.,\dots,\frac{1}{\sqrt{K}})$, while the eigenspace associated with eigenvalue $\lambda_k, k=2,\dots,K$ is of dimension $K-1$ and can be expressed as 
\[
E_{\lambda=1-\rho} =\{{\bf{u}}\in {R}^K: \sum_{k=1}^K u_k=0 \}.
\]
Acutally, there are infinitely many possible choices of the $K-1$ eigenvectors in $E_{\lambda=1-\rho}$  when $K \geq 3$, and hence infinitely many possible choice of $\mbox{PC}_k$, $k=2,\dots,K$.

The first eigenvalue $(K-1)\rho + 1$ is usually much larger than eigenvalue $1-\rho$ for relatively large $K$. Such a correlation structure is related to the spiked  population  co-variance model \citep{Johnstone2001}.  The principal angle between ${\bbeta}$ and ${\bf u}_1$ is 0 when ${ \bbeta}=(1,1,\dots,1)^Tc$ where $c$ is a  non-zero scalar, and is $90^o$ when $\sum_{k=1}^K{\beta_k}=0$. Therefore, $\mbox{PC}_1$ can best detect fully dense homogeneous signals, and its power decreases when the signals become sparser or in different directions. In addition, the power of $\mbox{PC}_1$  decreases when the correlation $\rho$ increases.  When  signals are fully dense and homogeneous, the WI test will also have good power, but the Wald, and VC tests might have low power. For example, in the simulation study, when $K=40$ and $\rho = 0.2$, the WI test has power of 0.82 to detect fully dense and homogeneous signals ${ \bbeta}=(1.4,1.4,\dots,1.4)^T$, but the Wald test has power of only 0.24 as shown in the setting M9 in Table \ref{power_result}.  However, if the signals are heterogeneous and $\sum_{k=1}^K{\beta_k}=0$ with at least one  $\beta_k$ nonzero,  $\bbeta$ is in the eigen-space $E_{\lambda=1-\rho}$ and can be detected by Wald and VC with good power but not WI.  

\subsection{Block Diagonal Exchangeable Correlation Matrices}
\label{sec_block_diag}
If  ${\bSigma}_{1}$ and ${\bSigma}_{3}$ are exchangeable with correlations $\rho_1$, $\rho_3$ respectively,   and the $K_1$ signal phenotyes are uncorrelated with the $K_0$ noise phenotypes, then the four unsorted eigenvalues of $\bf{\Sigma}$  and their algebraic multiplicities are
\begin{eqnarray*}
\lambda_1 &=& 1 + (K_1-1)\rho_1,\nu(\lambda_1)=1; \lambda_2 = 1-\rho_1,\nu(\lambda_2)= K_1 -1;\\
\lambda_3 &=& 1 +(K_0-1)\rho_3,\nu(\lambda_3)=1;  \lambda_4 = 1-\rho_3,\nu(\lambda_4)=K_0 -1.
\end{eqnarray*}
The signal phenotype eigenspaces are
\begin{eqnarray*}
E_{\lambda_1}&=& \{ {\bf{u}}\in {R}^K: {\bf{u}}^T= t (\underbrace{1/\sqrt{K_1},\dots,1/\sqrt{K_1},}_{K_1}\underbrace{0,\dots,0}_{K_0}), t\in {R}\},\nonumber\\
E_{\lambda_2}&=& \{ {\bf{u}}\in {R}^K: \sum_{k=1}^{K_1}u_k=0, u_{K_1+1}=\cdots=u_{K}=0 \},
\end{eqnarray*}
and the noise phenotype eigenspaces are
\begin{eqnarray*}
E_{\lambda_3}&=& \{ {\bf{u}}\in {R}^K: {\bf{u}}^T= t (\underbrace{0,\dots,0,}_{K_1}\underbrace{1/\sqrt{K_0},\dots,1/\sqrt{K_0}}_{K_0}), t\in {R}\},\nonumber\\
E_{\lambda_4}&=& \{ {\bf{u}}\in {R}^K: \sum_{k=K_1+1}^{K}u_k=0, u_{1}=\cdots=u_{K_1}=0 \}.
\end{eqnarray*}

Because the signal and noise phenotype eigenspaces are orthogonal to each other, thus those $K_0$ PCs in the noise eigenspace are all powerless to detect any signals. Therefore, we only need to focus on discussing the powers of the $K_1$ PCs in the signal phenotype eigenspaces. From Section \ref{sec_eigen_exch}, we know that the PC in  $E_{\lambda_1}$ can best detect homogeneous effects while the PCs in $E_{\lambda_2}$ can best detect heterogeneous effects. Actually, as long as some principal angle is  zero, then that particular PC with zero principal angle  in the signal eigenspace is as powerful as the Oracle test,  regardless of signal sparsity.

\subsection{Block Diagonal Correlation Matrices}
\label{sec_block_unstr}
We now consider more general situations where  ${\bSigma}_{1}$ and ${\bSigma}_{3}$ are unstructured.  By performing spectral decomposition on these two matrices, we have 
${\bSigma}_1 = {\bU}_1{\bLambda}_1 {\bU}_1^T $ and ${\bSigma}_3 = {\bU}_3{\bLambda}_3 {\bU}_3^T$, where ${\bLambda}_1$ and ${\bLambda}_3$ are diagonal matrices with diagonal elements the eigenvalues, ${\bU}_1$ and  ${\bU}_3$ are eigenvector matrices. If  the signal and noise phenotypes are uncorrelated, then we have

\[
\begin{pmatrix}
{\bSigma}_1& {\bf 0}\\
{\bf 0}& {\bSigma}_3
\end{pmatrix}
=
\begin{pmatrix}
{\bU}_1 & {\bf 0} \\
 {\bf 0}& {\bU}_3
\end{pmatrix}
\begin{pmatrix}
{\bLambda}_1 & {\bf 0} \\
{\bf 0} & {\bLambda}_3
\end{pmatrix}
\begin{pmatrix}
{\bU}_1 & {\bf 0} \\
 {\bf 0} & {\bU}_3
\end{pmatrix}^T.
\]

Therefore, the PCs from the signal eigenspace allocate zero loadings for the noise phenotypes. In other words, if signal and noise phenotypes are uncorrelated, then the PCs from the signal eigenspace are not contaminated by any noise phenotypes and thus one particular PC from the signal eigenspace can be as powerful as the Oracle test if its principal angle is zero, regardless of signal sparsity.

\section{Simulation Studies}
\label{sec_sim}

\subsection{Type I Error Rates}
Single PC tests and PCLC follow the standard normal distribution under the null so that their type I error rates are always well controlled, hence we omit their type I error results. Besides the Wald test,  we evaluate the sizes of the  proposed PC-based tests, including the $p$-value based tests PCMinP and PCFisher, and quadratic tests WI and VC, and the omnibus tests PCAQ and PCO, at the nominal levels $\alpha=0.05, 0.01, 0.001, 10^{-4}, 10^{-5}$, in view of the small significance levels that are of common interest in GWAS. For comparison purpose, we also included the $p$-value correction  method TATES \citep{vanderSluis2013} for comparison purpose which also only requires GWAS summary statistics.  We first consider a low-dimensional unstructured  covariance matrix ${ \bSigma}_{unK3}$  estimated from the global lipids data \citep{Teslovich2010} for high-density lipoprotein cholesterol (HDL), total cholesterol (TC) and triglycerides (TG),    
\begin{equation}
{\bSigma}_{unK3} =
\begin{bmatrix}
1.00 & 0.16 &-0.42  \\
0.16 & 1.00 & 0.38  \\
-0.42  & 0.38 &  1.00
\end{bmatrix}.\label{Cor_Mat_lipids}
\end{equation}
We also consider a high dimensional ($K=100$) unstructured covariance matrix ${ \bSigma}_{unK100}$ generated using the algorithm described in \citet{Marsaglia1984} and the actual matrix is provided in the supplementary excel file. We generated 10 millions of multivariate normal samples of dimensions $K=3$ and $K=100$ with mean zeros and covariance matrices equal to ${ \bSigma}_{unK3}$ and 
${ \bSigma}_{unK100}$ respectively. We found that the type I error rates of the PC based tests are well controlled  at those nominal levels as summarized in Table \ref{Table_size}. The type I error rates of the $p$-value correction method TATES are slightly inflated, in line with previous findings by \citet{He2013}.

\begin{table}[ht]
	\centering
	\def\~{\hphantom{0}}
	\begin{minipage}{180mm}
		\caption{Type I error rates estimated as the proportions of $p$-values less than significance level $\alpha$ in $10^7$ simulation replications under the nulls in both low and high dimensional settings. }
		\label{Table_size}
		\begin{tabular*}{\textwidth}{@{\extracolsep{\fill}} l@{}l@{}l@{}l@{}l@{}l@{}l@{}l@{}l@{}}
					\hline \\ [-6pt]
					& & \multicolumn{7}{c}{ Low dimensional setting: $K=3$, covariance matrix is ${ \bSigma}_{unK3}$ }\\ [1pt]
					\hline \\ [-6pt]
			$\alpha$ & PCMinP& PCFisher & WI & Wald &VC & PCAQ & PCO & TATES	\\ [-2pt]
			\hline\vspace{1pt}
			0.05 & 0.050  & 0.050 & 0.050 & 0.050 & 0.050  & 0.050  & 0.050 & 0.051  \\
			0.01 & 0.010 &	0.010 &	0.010 &	0.010&	0.010	&0.0099&	0.0099 & 0.011 \\
			0.001 & 0.001 &	0.00099  &0.001 &	0.001 &	0.001 &	0.001 &	0.00099 & 0.00106\\
			$10^{-4}$ & $9.75\times 10^{-5}$ &	$1.01\times 10^{-4}$  &$1.02\times 10^{-4}$ &	$1.01\times 10^{-4}$ &	$1.02\times 10^{-4}$ &	$9.81\times 10^{-5}$ &	$9.93\times 10^{-5}$ & $1.08 \times 10^{-4}$\\
			$10^{-5}$& $1.03 \times 10^{-5}$ &	$9.32\times 10^{-6}$	&	$8.33 \times 10^{-6}$& $8.61\times 10^{-6}$& $8.52\times 10^{-6}$ &	$8.47 \times 10^{-6}$ &	$9.71\times 10^{-6}$ & $1.05\times 10^{-5}$ \\
			\hline\\ [-6pt]
			& & \multicolumn{7}{c}{ High dimensional setting: $K=100$, covariance matrix is ${ \bSigma}_{unK100}$ }\\ [1pt]
			\hline\\[-6pt]
			$\alpha$ & PCMinP& PCFisher & WI & Wald &VC & PCAQ & PCO & TATES	\\ [-2pt]
			\hline\vspace{1pt}
			0.05 & 0.050  & 0.050 & 0.050 & 0.050 & 0.050  & 0.050  & 0.050 & 0.051  \\
			0.01 & 0.010 &	0.010 &	0.010 &	0.010&	0.010	&0.0099&	0.0099 & 0.011 \\
			0.001 & 0.001 &	0.00099  &0.001 &	0.001 &	0.001 &	0.001 &	0.00099 & 0.00106\\
			$10^{-4}$ & $9.65\times 10^{-5}$ &	$1.01\times 10^{-4}$  &$1.02\times 10^{-4}$ &	$1.01\times 10^{-4}$ &	$1.02\times 10^{-4}$ &	$9.86\times 10^{-5}$ &	$9.95\times 10^{-5}$ & $1.06 \times 10^{-4}$\\
			$10^{-5}$& $1.02 \times 10^{-5}$ &	$9.38\times 10^{-6}$	&	$8.46 \times 10^{-6}$& $8.78\times 10^{-6}$& $8.99\times 10^{-6}$ &	$9.37 \times 10^{-6}$ &	$9.78\times 10^{-6}$ & $1.05\times 10^{-5}$ \\
			\hline
		\end{tabular*}
	\end{minipage}
	\vspace*{-6pt}
\end{table}

\subsection{Power Comparisons of the PC Based Tests}
As shown in Figure \ref{PCA_rejectionBoundary}, different tests have different  rejection boundaries and the power of each test depends on both the mean vector and the covariance matrix of $\bf{Z}$.  We first provide empirical evidence using bivariate phenotypes to show that the powers of the PC based tests depend on  the direction of the true $\bbeta$ for a fixed between-phenoypte correlation matrix,  and no single test is  most powerful for all directions of $\bbeta$, while the omnibus tests are more robust. 

\begin{figure}[!t]
\centerline{\includegraphics[scale=0.4]{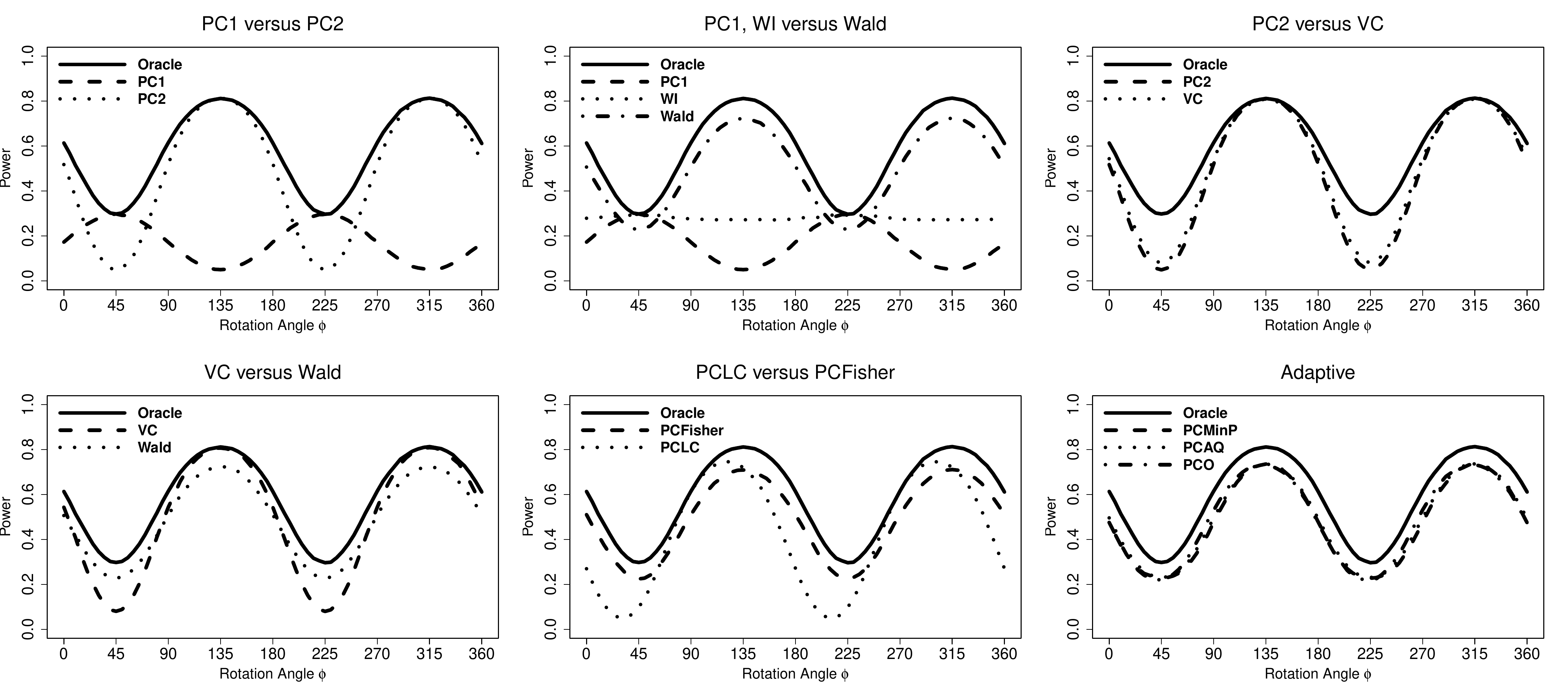}}
\caption[Rotation]{This figure shows the power curve of each PC test for alternatives ${\bbeta} = r(\cos(\phi),\sin(\phi))$ where $r=2$ and $\phi \in [0,360^o]$ are the polar coordinates. The bivariate correlation is $\rho =0.6$. This figure mirrors the rejection boundaries as shown in Figure \ref{PCA_rejectionBoundary}.}
\label{Rotation_power}
\end{figure}

Consider a bivariate standard normal $(Z_1,Z_2)^T$ with $\rho=0.6$ and mean  ${\bbeta} = (\beta_1,\beta_2)^T \neq \bf {0}$ under the alternative. Using the polar coordinate system, we can rewrite ${\bbeta} = r\{\cos(\phi),sin(\phi))\}$, where $r \geq 0$  and $\phi \in [0,360^o]$. For illustrative purpose, we set $r=2$. Then the power of each test is a function of $\phi$ only. We  divide the interval $[0,360^o]$ equally into 72 sub-intervals specified by 73 grid points, $\phi_b=0,5^o,\dots, 360^o, b=1,2,\dots,73$. For each $\phi_b$, we generated one million standard bivariate normal samples with $\rho=0.6$ and obtained one million $p$-values for each test. The power of each test for each $\phi_b$ is estimated by the proportion of $p$-values that are less than $\alpha =0.05$. By connecting the 73 power points of each test, we obtain the power function curves in Figure \ref{Rotation_power}. 

We found a periodic pattern of the power curves with period equal to $180^o$, and within each period, the power of each test is a function of $\phi$,  where $\phi$ specifies the direction of ${\bbeta}$ in $\mathcal{R}^2$.
 $\mbox{PC}_1$ is as powerful as the Oracle test when $\phi=45^o,225^o$ or equivalently when ${\boldsymbol \beta}=\pm(\sqrt{2},\sqrt{2})^T$, and consequently WI  is almost as powerful as the Oracle test in such settings. 
Likewise, $\mbox{PC}_2$ is as powerful as the Oracle test when $\phi=135^o,315^o$ or equivalently when ${ \bbeta}=\pm(-\sqrt{2},\sqrt{2})^T$, and VC is almost as powerful as the Oracle test (more powerful than Wald) in these settings. Wald and PCFisher have almost the same power curves. PCLC is as powerful as the Oracle test when $\phi=90^o,270^o$ or ${\bbeta}=\pm(0,2)^T$. The omnibus tests PCAQ and PCO  are never as powerful as the ideal Oracle test, but  are robust to the alternatives with little power loss compared to the Oracle test. When either the first or the last principal angle is zero,  PCAQ and PCO can be  more powerful than Wald.

\sloppy
In low dimensional settings where $K=3$ (M1-M5), we consider  an unstructured correlation matrix  ${\bSigma}_{unK3}$ given  in equation (\ref{Cor_Mat_lipids}) and the following five mean vectors:  ${\bbeta}_1=(-1.94, 1.58, 2.87)$, ${\bbeta}_2=(2.31, 2.62, 0.1)$, 
${\bbeta}_3=(0.99, -0.94, 1.18)$, ${\bbeta}_4=(0.94, 0.86, 1.92)$, ${\bbeta}_5=(0.79, 3.2, 0.16)$. We also consider $K=8$ and an unstructured correlation matrix ${\bSigma}_{unK8}$ given in Table \ref{MetS_Sigma} of Section \ref{sec_MetS}, and three mean vectors: ${\bbeta}_6=4.5{\bu}_1=(1.18, 0.69, 1.33, -2.39, 1.39, 2.66, 1.27, 0.52)$, ${\bbeta}_7=3.5{\bu}_4=(-0.8, 1.04, 0.74, 0.39, 1.69, -0.01, -0.67, -2.55)$,  and ${\bbeta}_8=2.5{\bu}_{8}=(0.08, -0.02, -0.11, 1.54, -0.65, 1.83, -0.08, -0.24)$, where ${\bu}_1, {\bu}_4, {\bu}_{8}$ denote the  first, fourth and eighth eigenvectors of ${\bSigma}_{unK8}$   respectively.

In high dimensional settings, we first consider  $K=40$ and an exchangeable correlation matrix  ${\bSigma}_{exK40}$ with off-diagonal correlation $\rho=0.2$, and a fully dense and  homogeneous signal vector ${\bbeta}_9=(1.4,1.4,\dots,1.4)$ as in setting M9. In setting M10, the correlation matrix ${\bSigma}_{bexK40}$ is block diagonal where the signal ($K_1=6$) and noise ($K_0=34$) blocks are exchangeable with correlations equal to $0.5$ and $0.2$ respectively, and the signal vector is sparse and ${\bbeta}_{10}=( 1.98, -1.51, -0.12, -0.12, -0.12, -0.12,0,\dots,0)$ with six nonzero elements and 34 zero elements. We also consider $K=100$ and an unstructured correlation matrix ${\bSigma}_{unK100}$ (provided in supplementary excel file) generated using the \citet{Marsaglia1984} algorithm. In settings M11 and M12, consider two sparse signal vectors $\bbeta_{11}=(0.03, 0.04, -0.02, 0.05, -0.04, 0.01, 0.02, 0.09, -0.13, -0.02, 0,\dots,0)$ and $\bbeta_{12}=(-0.17, 0.4, -0.05, 0.19, -0.68, 0.21, 0.3, -0.28, 0.29, -0.11, 0,\dots,0)$ that both contain ten non-zero elements and 90 zeros. In settings M13-M15, consider three dense signal vectors (provided in the supplementary excel file): $\bbeta_{13}$ contains 90 signals and ten noises, $\bbeta_{14}$ and $\bbeta_{15}$ are set to be proportional to the first and the eightieth eigenvectors of ${\bSigma}_{unK100}$ respectively.

For each setup, we generated $10^5$ multivariate normal samples with mean equal to $\bbeta$ and correlation matrix equal to $\bSigma$ and obtained $10^5$ $p$-values for each test. The empirical power was calculated as the proportion of $p$-values  less than $\alpha=0.05$. We summarize the power results in Table \ref{power_result}.  The results show that whenever $\theta_k=0$,  then $\mbox{PC}_k$ is as powerful as the Oracle test as shown in the low dimensional settings from M1 to M3. $\mbox{PC}_1$ requires a larger overall signal magnitude $|| \bbeta||$ to have comparable power as that of $\mbox{PC}_3$, simply because $PC_1$ has a larger variance. As expected, WI is more powerful than Wald and VC whenever $\mbox{PC}_1$ is the Oracle test,  while VC is more powerful than WI and Wald whenever the last PC is the Oracle test. 
PCLC is as powerful as the Oracle test as shown in the setting M4 where the three principal angles are equal to each other, and PCLC is powerless in the setting M5 where the signal vector $ \bbeta$ is parallel to the rejection boundary of PCLC.  The TATES method can have comparable power to PCO in settings M1, M2 and M5, but it can perform poorly in settings M3 and M4. By contrast, the PCO always has good power in all those five settings M1-M5. In M6, the $\mbox{PC}_1$ test attains the Oracle power, and hence WI and PCMinP are both more powerful than the Wald test, the adaptive omnibus tests PCAQ and PCO also outperform Wald. In setting M7 where we set the fourth principal angle to be zero, and thus the Wald test outperform WI and VC in this setting, but is still less powerful than PCMinP. PCO outperforms Wald because  PCMinP is one of its combining component. In M8 where the last principal angle is set to be zero,  the last PC is the Oracle test and VC is  more powerful than Wald. PCAQ and PCO also outperform Wald. The TATES method  has comparable power to PCO in M6 but performs poorly than PCO in M7-M8.    
 
In high dimensional setting M9, $\mbox{PC}_1$ has the Oracle power to detect fully dense homogeneous signals simply because the first principal angle is zero,  while both  PCFisher and Wald have very low power in this setting. The TATES method is less powerful than the PCO in M9. 
In M10 where the signals are sparse, we additionally considered the MinP test defined as the minimum $p$-values across all the $K$ original Z-testing statistics as in \citep{Conneely07}, which is designed to detect sparse signals.
The power of MinP (not reported in Table \ref{power_result}) is 0.20, smaller than the powers of PCFisher, Wald, VC, PCAQ and PCO. This surprising
 result demonstrates that PC based tests can outperform MinP  for detecting sparse signals by leveraging on the between-phenotype correlation structures. The TATES method also has very low power in M10.  In high dimensional settings M11 and M12 where the signals are sparse, the MinP and the TATES methods are almost powerless while PCO has very good power to detect these two sparse signals.  The Wald test also performs poorly in these two settings. For dense signals in settings M13 and M15, the TATES method is almost powerless  and the Wald test also has very low power, while the PCO test still has good power. In setting M14 where we set the first principal angle to be zero, $\mbox{PC}_1$ is the Oracle test and the WI test is thus very powerful, while the Wald test has very low power.  The TATES method has comparable power to the PCO test in M14. We found that the TATES method has similar performance to the WI test, which can be explained by the similarities between the rejection boundaries of these two tests. The rejection boundary of the TATES method is provided in Figure S1 in Section S1 of the Supplementary Materials.

\begin{table}[ht]
 \centering
 \def\~{\hphantom{0}}
\caption{Parameter configurations for power comparison in  simulation studies.  The numbers in the three columns correspond to principal angles $\theta_1,\theta_2, \theta_K$ are in the unit of degree. The last column represents the power of the Oracle test.}
\label{power_setup}
  \begin{tabular*}{\textwidth}{@{\extracolsep{\fill}} l@{}l@{}l@{}c@{}l@{}l@{}l@{}l@{}l@{}l@{}l@{}}
\hline\\ [-10pt]
{Setup} & $K$ & $ \bbeta$ & $\bSigma$ & $||\boldsymbol \beta||$ & $\theta_1$  & $\theta_2$& $\theta_K$& Oracle  \\ [-2pt]
\hline
M1 & 3 & $\bbeta_1$ & ${\bSigma}_{unK3}$& 3.8    & 0      & 90     & 90         & 0.85   \\
M2 & 3 & $\bbeta_2$  & ${\bSigma}_{unK3}$& 3.5  & 90     & 0          & 90     & 0.89   \\
M3 & 3 & $\bbeta_3$ &${\bSigma}_{unK3}$  &1.8  & 90     & 90         & 0    & 0.85   \\
M4 & 3 & $\bbeta_4$  &${\bSigma}_{unK3}$ & 2.3  & 54.7     & 54.7        & 54.7      & 0.79   \\
M5 & 3 & $\bbeta_5$   & ${\bSigma}_{unK3}$ &3.3  & 71.2  & 27.5     & 109.7   & 0.91   \\
M6 & 8 & $\bbeta_6$ &${\bSigma}_{unK8}$  &4.5  & 0     & 90         & 90    & 0.93   \\
M7 & 8 & $\bbeta_7$  &${\bSigma}_{unK8}$ & 3.5  & 90     & 90        & 90      & 0.94   \\
M8 & 8 & $\bbeta_8$   & ${\bSigma}_{unK8}$ &2.5 &90  & 90  & 0      & 0.92   \\
M9 & 40 &  $\bbeta_9$  & ${\bSigma}_{exK40}$& 8.85 & 0	& 90 &	90   & 	0.83\\
M10 & 40 & $\bbeta_{10}$  & ${\bSigma}_{bexK40}$ &2.5 & 90  & 90  & 0   &  0.94\\
M11 & 100 & $\bbeta_{11}$  &${\bSigma}_{unK100}$ &0.18 & 90	& 90.3 &	3.6   &0.92 	\\
M12 &  100& $\bbeta_{12}$ &  ${\bSigma}_{unK100}$ & 1.0  & 90  &90    &90 &0.97 \\
M13 &  100& $\bbeta_{13}$ & ${\bSigma}_{unK100}$&3.38  & 90	&90  &	90  &0.97 	\\
M14 & 100 & $\bbeta_{14}$ &${\bSigma}_{unK100}$ & 15.0& 0  &90   & 90    &0.88  \\
M15 & 100 & $\bbeta_{15}$ & ${\bSigma}_{unK100}$& 2.0 & 90  & 90  & 90    &  0.97\\
\hline
\end{tabular*}
\vspace*{-6pt}
\end{table}

\begin{table}[!htbp]
 \centering
 \def\~{\hphantom{0}}
  \caption{Powers estimated as the proportions of $p$-values less than significance level $\alpha =0.05$ in $10^5$  replications under various alternatives. The setups are described in Table \ref{power_setup}.}
\label{power_result}
\centerline{
  \begin{tabular*}{\textwidth}{@{\extracolsep{\fill}}  c@{}  c@{} c@{} c@{}c@{}c@{}c@{}c@{}c@{}c@{}c@{}c@{}c@{}}
\toprule
& & \multicolumn{10}{c}{ Low dimensional setting: $K=3$ and $K=10$}  &\\ [1pt]
			\hline\\[-6pt]
{Setup}  & $\mbox{PC}_1$ & $\mbox{PC}_2$ & $\mbox{PC}_K$ &  PCMinP & PCFisher & PCLC & WI & Wald &VC & PCAQ & PCO & TATES	\\ [-2pt]
\hline
M1   & \textbf{0.85} & 0.05 & 0.05 & 0.76   & 0.72     & 0.23 & \textbf{0.81} & 0.74 & 0.31 & 0.76 & 0.75 &0.77  \\
M2 & 0.05 & \textbf{0.89} & 0.05 & 0.80   & 0.75     & 0.30 & 0.79 & 0.77 & 0.44 & 0.73 & 0.77  &0.77 \\
M3 & 0.05 & 0.05 & \textbf{0.85} & 0.74   & 0.70     & 0.67 & 0.23 & 0.72 & \textbf{0.84} & 0.81 & 0.77&  0.28 \\
M4 & 0.17 & 0.23 & 0.59 & 0.53   & 0.63     & \textbf{0.79} & 0.42 & 0.65 & 0.68 & 0.65 & 0.72 & 0.43 \\
M5 & 0.14 & 0.75 & 0.45 & 0.75   & 0.82     & \textbf{0.05} & 0.73 & 0.82 & 0.71 & 0.79 & 0.78  &0.78 \\
M6&0.93 & 0.05 & 0.05 & 0.76 & 0.61 & 0.14 & 0.84 & 0.67 & 0.32 & 0.80 & 0.75 & 0.71 \\
M7&0.05 & 0.05 & 0.05 & 0.81 & 0.66 & 0.23 & 0.67 & 0.72 & 0.68 & 0.66 & 0.75 & 0.59 \\
M8&0.05 & 0.05 & 0.92 & 0.76 & 0.61 & 0.34 & 0.33 & 0.67 & 0.83 & 0.78 & 0.74 & 0.34\\
\midrule
			& & \multicolumn{10}{c}{ High dimensional setting: $K=40$ and $K=100$} &\\ [1pt]
			\hline\\[-6pt]
			{Setup}  & $PC_1$ & $PC_2$ & $PC_K$ &  PCMinP & PCFisher & PCLC & WI & Wald &VC & PCAQ & PCO & TATES	\\ [-2pt]
\hline
M9     & \textbf{0.83} & 0.05 & 0.05& 	0.42 &	\textbf{0.21} &	0.06 &	\textbf{0.82} &	\textbf{0.24}& 	0.06& 0.75&0.68 &0.59 \\
M10  & 0.05 & 0.05  & 0.94 &   0.64 & 0.30 & 0.12 & 0.10 & 0.36 & 0.55 & 0.45 &0.56 &\textbf{0.20} \\
M11  & 0.05 & 0.05 & 0.92 & 0.53 & 0.18 & 0.66 & 0.05 & 0.22 & 0.92 & 0.85 & 0.83 & \textbf{0.05} \\
M12  & 0.05 & 0.05 & 0.05 & 0.75 & 0.24 & 0.09 & 0.05 & 0.31 & 0.07 & \textbf{0.19} & 0.63 & \textbf{0.06} \\
M13 & 0.06 & 0.05 & 0.05 & 0.73 & 0.23 & 0.06 & 0.07 & 0.28 & 0.05 & \textbf{0.19} & 0.61 & \textbf{0.09} \\
M14 & 0.88 & 0.05 & 0.05 & 0.42 & 0.15 & 0.04 & 0.86 & \textbf{0.18} & 0.05 & 0.77 & 0.71 & 0.70 \\
M15 & 0.05 & 0.05 & 0.07 & 0.72 & 0.24 & 0.06 & 0.06 & 0.30 & 0.07 & 0.20 & 0.58 & \textbf{0.06}\\
\bottomrule
\end{tabular*}
}
\vspace*{-6pt}
\end{table}

\section{Joint Analysis of Multiple Metabolic Syndrome Related Phenotypes }
\label{sec_MetS}
 We are interested in detecting the genetic associations between  individual SNPs and multiple phenotypes of metabolic syndrome using the GWAS summary statistics of the MetS-related phenotypes from the four international consortia described in the Introduction Section.     The GWAS summary statistics from these four consortia are publicly available. The website links for those data sets are provided in Section S2 in the Supplementary Materials. However, the individual level phenotype and genotype data are not directly accessible. Hence,  any multiple phenotype analysis method that requires individual level data cannot be applied.   The single-trait GWAS analysis performed by the four international consortia might miss susceptible SNPs that are associated with MetS, even with very large sample sizes,  because the genetic effects of common variants are usually small. 

To increase analysis power for identifying additional SNPs associated with MetS,  we applied the proposed  PC based testing procedures and the TATES method \citep{vanderSluis2013} to the GWAS summary statistics data by jointly analyzing the eight MetS-related traits described in the Introduction Section. They include Body Mass Index (BMI), Fasting Glucose (FG), Fasting Insulin (FI), High-Density Lipoprotein cholesterol (HDL), Low-Density Lipoprotein cholesterol (LDL), triglycerides (TG),   Waist-hip-ratio (WHR), and Systolic Blood Pressure (SBP).     We first merged these four GWAS summary statistics data sets using the common 1,999,568 SNPs shared by the four datasets.  We then performed our proposed PC based tests using these univariate Z-scores,  and also applied the TATES method on the univariate $p$-values. The correlation matrix  $\bSigma$ among these MetS related traits was estimated by the sample correlation matrix across approximately independent SNPs after LD pruning \citep{Zhu2015,liu2017}, and is provided in Table  \ref{MetS_Sigma}. 

\begin{table}[ht]
\centering
\caption{The correlation matrix $\bSigma$ of single-trait GWAS $Z$-scores estimated using the MetS GWAS summary statistics data sets. The eigen-values of $\bSigma$ are: 1.75, 1.26, 0.99, 0.95, 0.94, 0.82, 0.75, 0.54. }
  \vspace{0.1in}
  \label{MetS_Sigma}
\begin{tabular}{l|cccccccc}
$\bSigma$ & BMI & FG & FI & HDL & LDL & TG  & WHR & SBP \\
\hline
BMI&1     & -0.02  & -0.04      & -0.2  & 0.05  & 0.16  & -0.01 & -0.03 \\
FG&-0.02 & 1      & 0.2        & -0.02 & 0.01  & 0.05  & 0.03  & 0.08  \\
FI&-0.04 & 0.2    & 1          & -0.11 & 0.03  & 0.15  & 0.12  & 0.08  \\
HDL&-0.2  & -0.02  & -0.11      & 1     & -0.09 & -0.42 & -0.11 & 0     \\
LDL&0.05  & 0.01   & 0.03       & -0.09 & 1     & 0.24  & 0.06  & 0     \\
TG&0.16  & 0.05   & 0.15       & -0.42 & 0.24  & 1     & 0.15  & 0.07  \\
WHR&-0.01 & 0.03   & 0.12       & -0.11 & 0.06  & 0.15  & 1     & 0.06  \\
SBP&-0.03 & 0.08   & 0.08       & 0     & 0     & 0.07  & 0.06  & 1  
\end{tabular}
\end{table}

 The QQ plots for the GWAS analysis using the PCA-based methods and the TATES method are provided in Section S2 of the Supplementary Materials and show the proposed methods had good genomic control.   If a SNP was associated with any of the eight MetS related phenotypes, then it should have been reported previously in the single trait analysis published in the literature. Therefore, the newly detected SNPs are those SNPs that were detected by multiple phenotype analysis but were missed by previous single-trait analysis performed by the four international consortia. In other words, the $p$-values of those newly detected SNPs are not genome-wide significant in any of the eight single-trait GWAS studies.  Since the identified SNPs might be in Linkage Disequilibrium (LD) with each other,  we performed LD pruning  to obtain almost independent SNPs  using the LD threshold $r^2 < 0.01$ within each 500kb region by PLINK \citep{Purcell2007}. After LD pruning, we greatly reduced the numbers of newly detected significant SNPs, indicating that many newly detected SNPs are  in LD with each other.  The numbers of new SNPs detected by each test are summarized in Table \ref{MetS_snp_num}. 

\begin{figure}
\centering
    \begin{subfigure}{.2\textwidth}
        \centering
        \includegraphics[width=\linewidth]{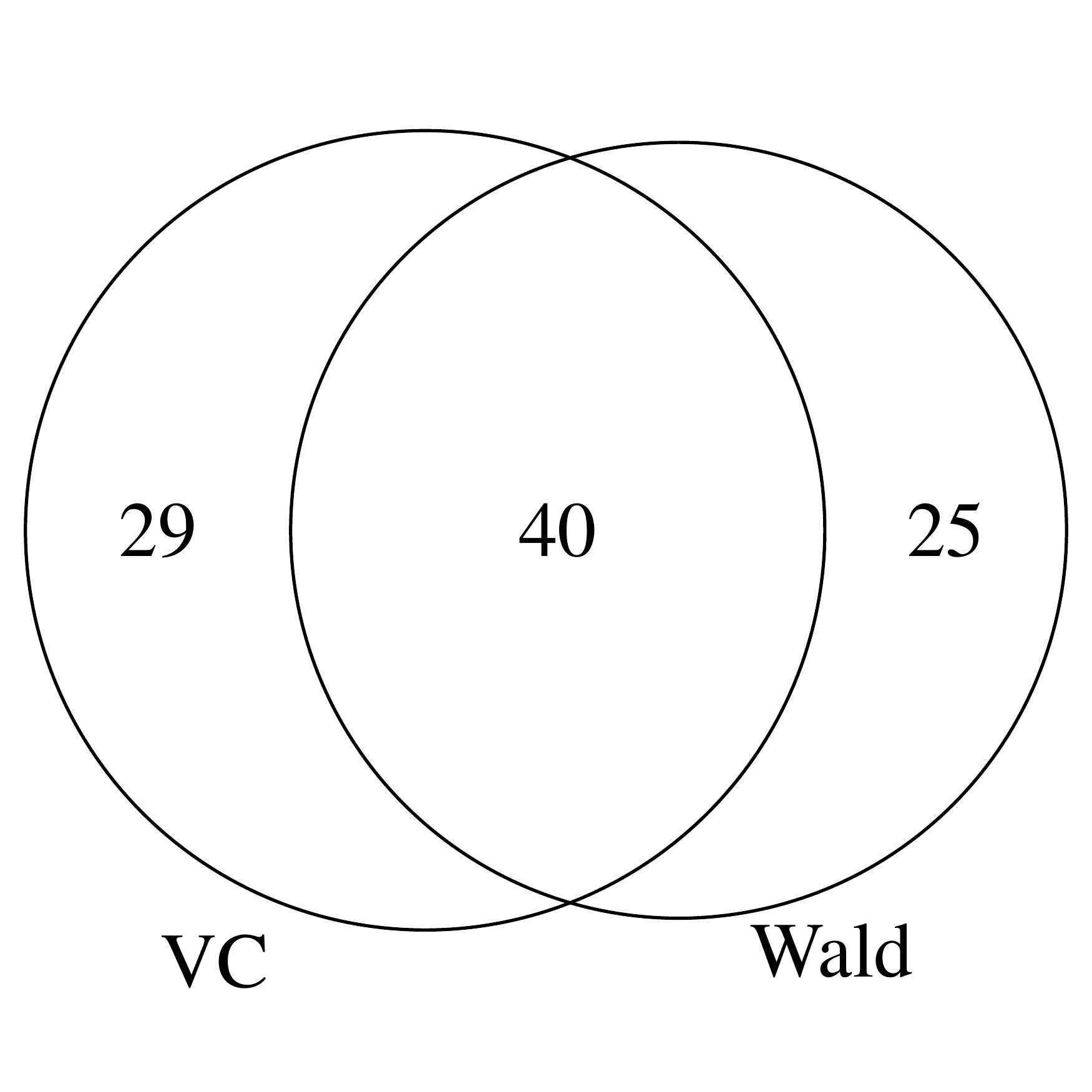}
        \caption{VC vs Wald}\label{fig:fig_a}
    \end{subfigure} %
    \begin{subfigure}{.2\textwidth}
        \centering
        \includegraphics[width=\linewidth]{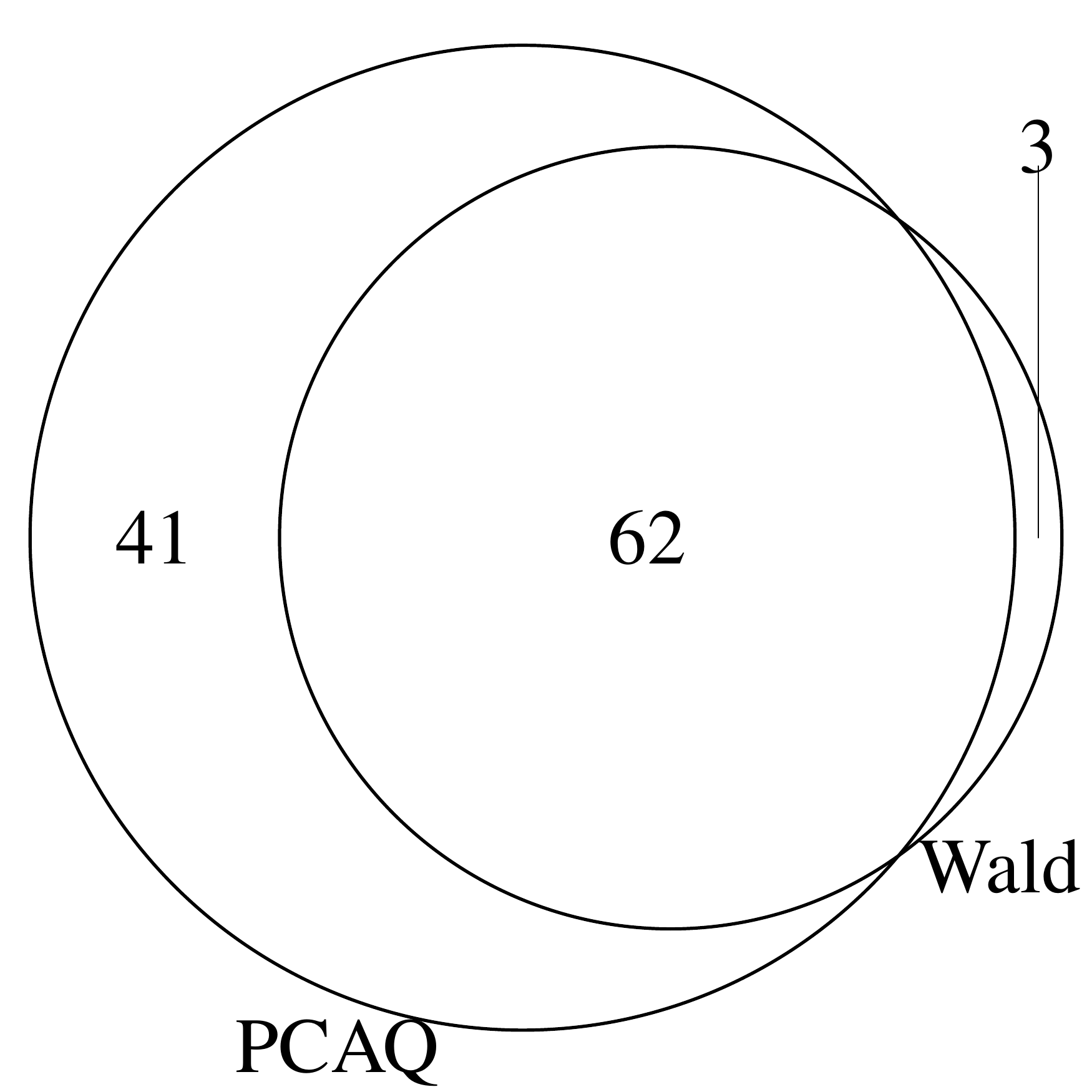}
        \caption{PCAQ vs Wald}\label{fig:fig_b}
    \end{subfigure} %
    \begin{subfigure}{0.2\textwidth}
        \centering
        \includegraphics[width=\linewidth]{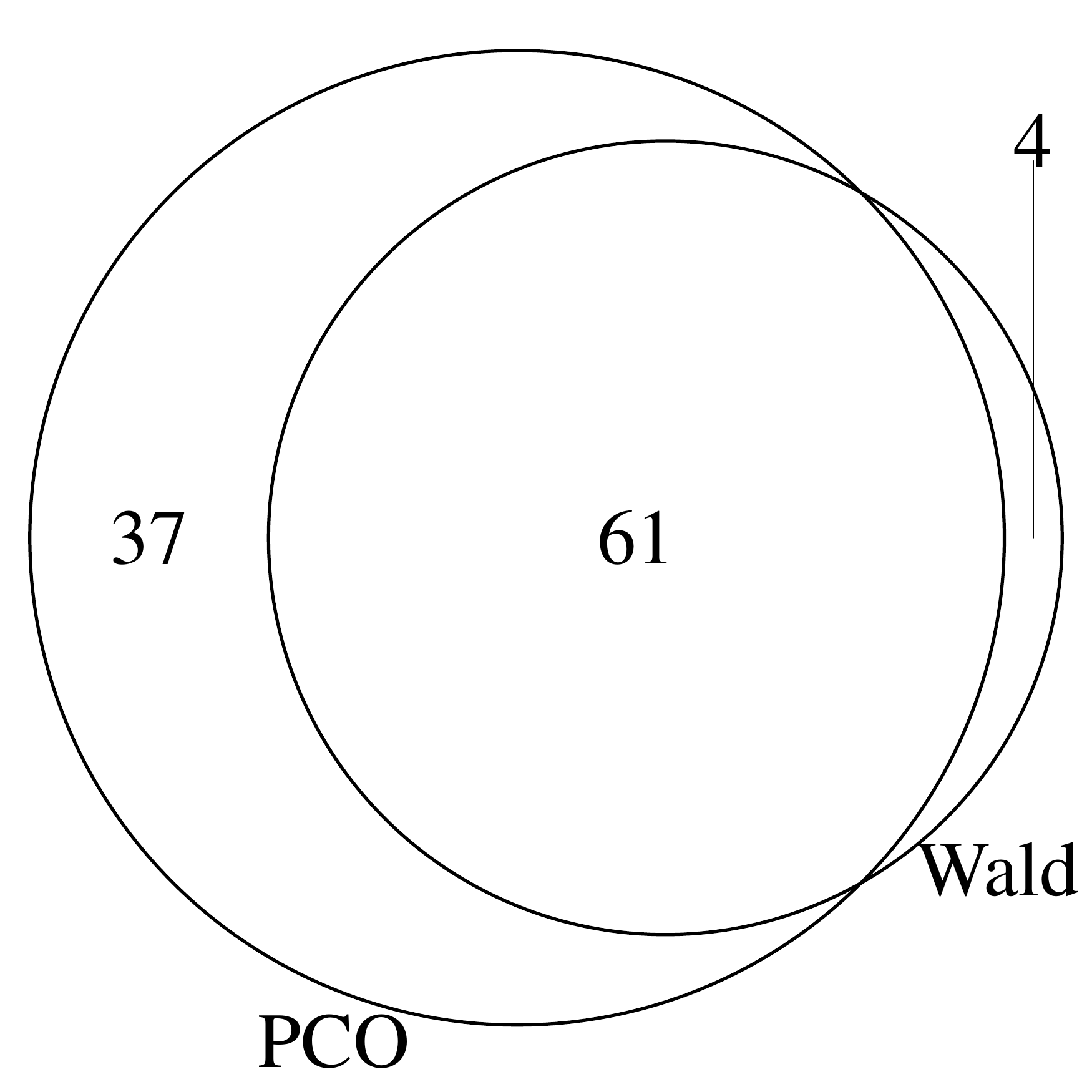}
        \caption{PCO vs Wald}\label{fig:fig_c}
    \end{subfigure}
      \begin{subfigure}{0.2\textwidth}
        \centering
        \includegraphics[width=\linewidth]{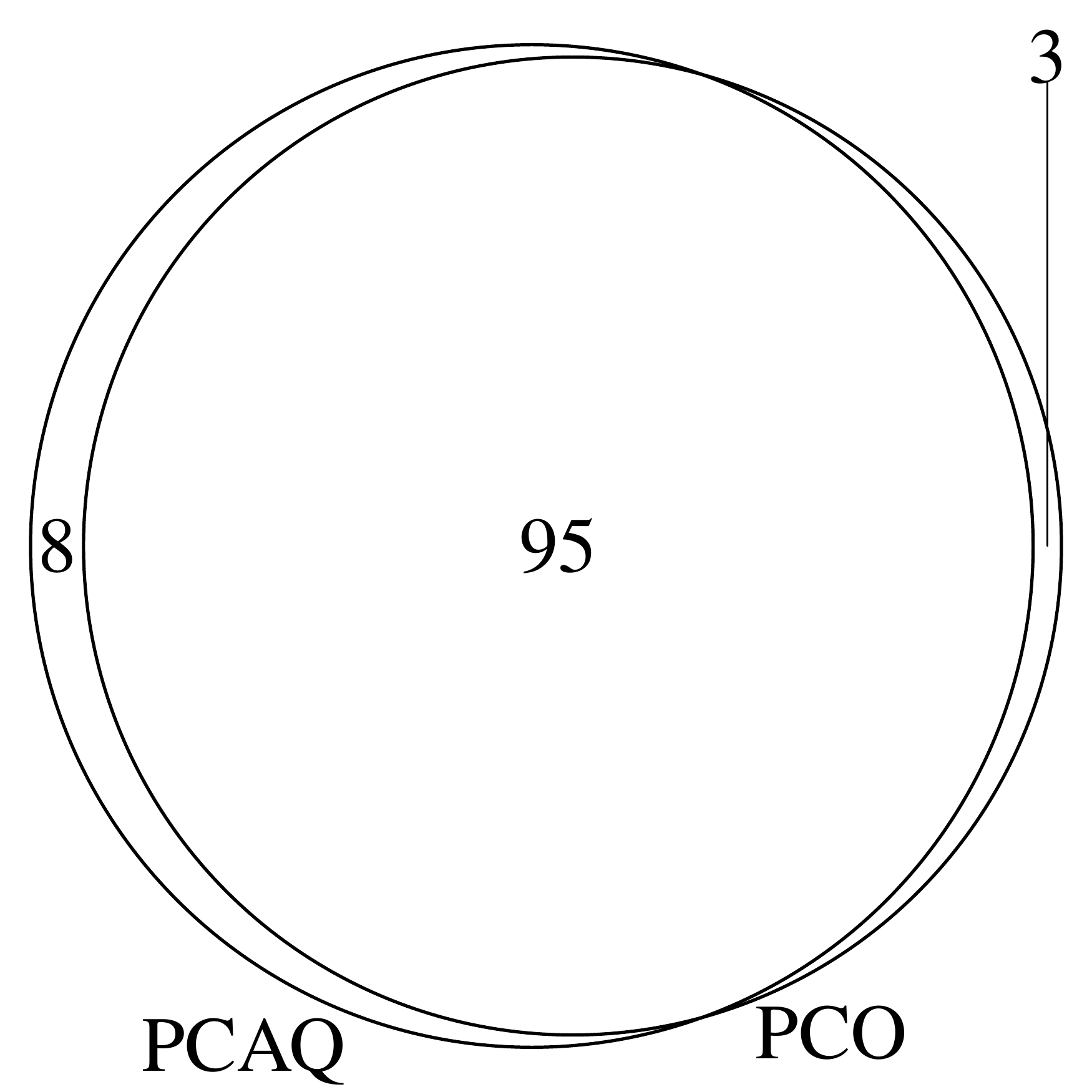}
        \caption{PCAQ vs PCO}\label{fig:fig_d}
    \end{subfigure}
\caption{{Venn Diagrams for  the overlapping SNPs detected by Wald, VC, PCAQ and PCO tests after LD pruning.}}
\label{MetS_M8_Venn}
\end{figure}

In what follows, we only report and discuss the identified SNPs after LD pruning. The last PC detected  26  SNPs after LD pruning, more than the other seven PCs. As expected, VC detected more SNPs than Wald and WI. PCAQ, which combines WI, Wald and VC, detected 103 SNPs, while PCO detected 98 SNPs. PCO detected slightly fewer SNPs than PCAQ because PCLC, PCFisher and PCMinP did not contribute more new SNPs in addition to the WI, Wald and VC tests.  Note by combining more tests, PCO pays  a higher price than PCAQ in the $p$-value adjustment. This is because PCO takes the smallest p-value of the six tests as the test statistic, while PCAQ only uses the smallest p-value  of the three tests (WI, Wald and VC) as the test statistic.  Many SNPs can be detected by more than one test. For example, 95 SNPs can be detected by both PCAQ and PCO as shown in  Figure \ref{MetS_M8_Venn}. It can also be seen from Figure \ref{MetS_M8_Venn} that VC and Wald detected 40 SNPs in common, while VC detected 29 SNPs that Wald failed to detect, and Wald detected 25 SNPs that VC failed to detect. This illustrates that each test can be more powerful than the others in some scenarios because of its admissibility property. It should be noted that our proposed adaptive omnibus tests PCAQ and PCO detected more SNPs than non-adaptive tests, demonstrating their robust performance. 

The TATES method only detected three new SNPs missed by the original single-trait GWAS studies. They are rs9600212 ($p = 1.08\times 10^{-13}$), rs9592962 ($p = 1.45\times 10^{-12}$), rs9592961 ($p = 7.12\times 10^{-12}$), within a 2kb intronic region of gene KLF12 on chromosome 13  before LD pruning. The most significant SNP rs9600212  was retained after LD pruning. These three SNPs  were also detected by PCFisher, WI, Wald, VC, PCAQ and PCO.   Note that KLF12  was found to be associated with the duration of the  Q, R, and S waves (QRS duration), which measures the duration of ventricular muscle depolarization seen on a typical electrocardiogram  and hence might play a role in affecting heart functions \citep{sotoodehnia2010common}. 

Note that it is possible that a SNP can be detected by single-trait analysis but might not be detected by multiple trait analysis using PC based  tests or the TATES method.  For example, SNP rs6129779 on chromosome 20 was found to be associated with LDL ($p=4.04\times 10^{-9}$) and has been reported by the GLGC \citep{Willer2013}, but it was not detected using the proposed PC based tests or the TATES method.  This is because there is only one weak signal mixed with seven noises, and thus joint analysis using either PC based tests or the TATES method might not be able to detect this rare and weak signal  with sufficient power. 

\begin{table}[ht]
 \centering
  \caption{The numbers of newly detected SNPs (not reported by the original GWAS studies) that reached the genome-wide significance at $\alpha=5\times 10^{-8}$  by joint analysis of the  eight MetS-related traits (BMI, FG, FI, HDL, LDL, TG, WHR, SBP) using  the proposed PC based tests and the TATES method before and after LD pruning.    }
\label{MetS_snp_num}
\begin{tabular}{p{2cm}p{1.2cm}p{1.2cm}p{1.2cm}p{1.2cm}p{1.2cm}p{1.2cm}p{1.2cm}p{1.2cm}p{1.2cm}}
\toprule
LD Pruning   & PC1 & PC2 & PC3   & PC4 & PC5 & PC6 & PC7  & PC8 &    \\
\hline
Before   &23&	7&	16	&6 &	29	&32	&19	&108 &     \\
After    &7	& 2	&3&	1	&5	&4&	1	&26&	 \\
\hline\\[-10pt]
LD Pruning    & PCMinP & PCFisher & PCLC & WI  & Wald & VC  & PCAQ & PCO & TATES   \\
\hline
Before     & 123	&404 &	7	&210	&458	&476	&682	&581	&3   \\
After    &25	&60 &	3	&42	&65	&69	&103	&98	&1     \\
\bottomrule
\end{tabular}
\end{table}

We now take a subset of the newly detected SNPs  presented in Table \ref{Table_MetS-pvalues} to illustrate the differences and connections of our proposed PC based tests.  For the ten SNPs in Table \ref{Table_MetS-pvalues}, none of their phenotype-specific $p$-values reached the genome-wide significance threshold, so
those ten SNPs were not identified by the original single-trait analysis performed by the four international consortia. We also estimated the eight empirical principal angles (provided in the supplementary excel file) for each  of the ten SNPs by calculating the angles between the Z-score vector and the eigenvectors of the correlation matrix in Table \ref{MetS_Sigma} to help illustrate the concept of principal angle in this real data example. For SNP rs355838, its first principal angle was estimated as $41.8^o$ and the $p$-value of PC1 was $1.53\times 10^{-9}$, and all the other seven principal angles for this SNP were more closer to $90^o$ and the $p$-values of all the other seven PCs were not genome-wide significant. Intuitively, this means that the genetic effect vector of SNP rs355838 is more similar to the first PC direction and less similar to the other PC directions. Using the first PC will more likely detect this association signal.   As a result, the $p$-value of WI, which has a similar performance to PC1,  for detecting SNP rs355838 is more significant than those of PCFisher, Wald and VC. Biologically, SNP rs355838 is located in an intronic region of gene COBLL1, which was reported as  a pleiotropic gene  that was  associated with metabolic syndrome and inflammation by \citep{kraja2014}. Specifically, this SNP was found to be associated with at least one metabolic trait and one inflammatory marker. 

The last principal angle of SNP rs8321 was estimated to be $38^o$, and all the other seven principal angles for this SNP were more closer to $90^o$.  The $p$-value of the last PC was $1.71\times 10^{-11}$, and hence VC was more significant than Wald, WI, PCFisher, PCLC and PCMinP. Another example is SNP rs308971 whose first principal angle was estimated to be $29.26^o$ and the $p$-value of the first PC was $4.39\times 10^{-8}$, while the $p$-values of all the other seven PCs were not genome-wide significant, because their principal angles are closer to $90^o$. For this SNP  rs308971, the $p$-value of WI test was   $2.48\times 10^{-10}$ while the Wald and VC tests were not even genome-wide significant.  This gene SYN2 was found to be related to type 2 diabetes (T2D) \citep{zeggini2008meta}. 

As for SNP rs9394279, the $p$-value of PCLC was $1.4 \times 10^{-8}$ while the $p$-values of WI, Wald, VC and PCAQ were not genome-wide significant. The PCO test which contains PCLC as one combining component has $p$-value $2.98 \times 10^{-8}$. This demonstrates that PCO which combines PCLC, PCMinP, PCFisher, WI, Wald and VC  all together is more robust than any  of the individual components, and is also more robust than PCAQ which only combines three tests: WI, Wald and VC. We can also see from Table \ref{Table_MetS-pvalues} that whenever any of WI, Wald and VC is significant, then the p-value of PCAQ is slightly more significant than PCO. This is because PCLC, PCMinP and PCFisher contribute little or none information in addition to WI, Wald and VC when the latter three tests can already capture the signal, and in this case PCAQ will perform slightly better than PCO. However, as in case of SNP rs9394279, WI, Wald, VC and hence PCAQ failed to detect this signal, but PCO was able to detect it. This is because PCLC can capture this signal. Those identified new SNPs provide potential candidates for future functional studies to better understand their biological roles in the etiology of metabolic syndrome. 

\begin{table}[ht]
\centering
\caption{ P-values of a selected subset of new SNPs detected by PC based tests. The $p$-values of PCAQ and PCO for the first SNP rs355838   are reported  as $<10^{-15}$ due to the numerical precision limits of the R package {\it mvtnorm}.  CHR represents chromosome number.}
\begin{adjustbox}{width=1\textwidth}
\label{Table_MetS-pvalues}
\begin{tabular}{llllllllllll}
\toprule
rsID       & CHR & Gene    & FG   & FI & HDL    & LDL     & TG        & WHR    & BMI   &  SBP & \\
\hline 
rs355838   & 2  & COBLL1  & 1.78E-01 & 1.56E-07 & 4.10E-07 & 3.05E-05 & 1.21E-04 & 1.70E-07 & 7.80E-06 & 1.13E-01 \\
rs8321     & 6  & ZNRD1   & 1.12E-01 & 2.96E-01 & 1.69E-05 & 1.91E-01 & 1.05E-06 & 1.10E-01 & 3.87E-01 & 3.19E-01 \\
rs5754352  & 22 & UBE2L3  & 5.21E-02 & 9.09E-01 & 6.57E-08 & 3.33E-02 & 9.11E-03 & 5.70E-01 & 5.93E-04 & 2.88E-01 \\
rs308971   & 3  & SYN2    & 5.25E-03 & 2.78E-06 & 3.27E-03 & 2.36E-02 & 3.51E-05 & 1.50E-04 & 5.54E-01 & 1.29E-02 \\
rs2269928  & 11 & C11orf9 & 4.64E-01 & 4.76E-01 & 7.63E-02 & 9.59E-08 & 1.32E-06 & 3.20E-01 & 7.87E-01 & 2.50E-01 \\
rs10744777 & 12 & ALDH2   & 9.99E-01 & 8.40E-01 & 1.46E-02 & 1.56E-07 & 2.18E-02 & 8.90E-01 & 1.18E-01 & 6.24E-06 \\
rs11717195 & 3  & ADCY5   & 2.68E-07 & 1.20E-01 & 4.77E-04 & 3.77E-02 & 4.52E-01 & 2.60E-01 & 1.34E-04 & 5.05E-01 \\
rs6485702  & 11 & LRP4    & 4.60E-02 & 8.54E-02 & 6.88E-08 & 4.68E-04 & 2.40E-07 & 2.10E-01 & 4.72E-01 & 2.99E-01 \\
rs6810027  & 3  & NISCH   & 4.11E-01 & 1.06E-02 & 1.07E-06 & 1.67E-02 & 3.09E-02 & 1.20E-07 & 8.68E-02 & 2.37E-01 \\
rs9394279  & 6  &intergenic   & 5.02E-01 & 2.62E-03 & 2.92E-05 & 2.96E-02 & 6.46E-01 & 1.60E-01 & 1.56E-01 & 2.27E-01\\
\midrule
SNP        & CHR & Gene     & PCLC    & PCFisher & PCMinP  & WI       & Wald    & VC     & TATES   & PCAQ   & 
PCO    \\
\hline
rs355838   & 2  & COBLL1  & 1.42E-03 & 1.10E-16 & 1.23E-08 & 9.62E-18 & 1.28E-16 & 5.14E-11 & 6.15E-06 & $<$E-15 & $<$E-15 \\
rs8321     & 6  & ZNRD1   & 3.05E-06 & 1.25E-09 & 1.37E-10 & 1.28E-05 & 1.50E-10 & 9.30E-14 & 2.78E-04 & 9.44E-14 & 1.82E-13 \\
rs5754352  & 22 & UBE2L3  & 3.69E-05 & 2.04E-09 & 8.74E-07 & 4.28E-06 & 1.54E-09 & 7.13E-11 & 2.69E-05 & 7.13E-11 & 1.40E-10 \\
rs308971   & 3  & SYN2    & 1.73E-02 & 2.32E-06 & 3.51E-07 & 2.48E-10 & 2.32E-07 & 1.32E-03 & 7.29E-05 & 2.48E-10 & 4.92E-10 \\
rs2269928  & 11 & C11orf9 & 9.57E-02 & 5.71E-09 & 3.71E-05 & 7.67E-06 & 4.61E-09 & 5.07E-10 & 6.98E-06 & 5.38E-10 & 1.03E-09 \\
rs10744777 & 12 & ALDH2   & 1.11E-02 & 6.14E-10 & 1.43E-06 & 1.90E-07 & 6.63E-10 & 7.54E-09 & 2.24E-05 & 6.73E-10 & 1.22E-09 \\
rs11717195 & 3  & ADCY5   & 1.78E-06 & 1.18E-09 & 8.52E-04 & 1.69E-07 & 2.18E-09 & 4.71E-08 & 8.51E-06 & 2.18E-09 & 2.33E-09 \\
rs6485702  & 11 & LRP4    & 2.15E-01 & 7.79E-07 & 8.47E-07 & 2.07E-09 & 3.65E-07 & 5.21E-04 & 5.01E-06 & 2.16E-09 & 5.35E-09 \\
rs6810027  & 3  & NISCH   & 2.56E-03 & 1.69E-08 & 2.34E-05 & 4.74E-09 & 1.72E-08 & 1.70E-06 & 4.21E-06 & 4.94E-09 & 9.58E-09 \\
rs9394279  & 6  & 		intergenic     & 1.40E-08 & 2.36E-06 & 4.59E-03 & 1.34E-04 & 3.18E-06 & 1.71E-06 & 4.23E-04 & 2.48E-06 & 2.98E-08\\
\bottomrule
\end{tabular}
\end{adjustbox}
\end{table}

\section{Discussion}
\label{sec_discussion}
In this paper, we proposed a series of principal component based testing procedures to detect genetic associations between a SNP and multiple phenotypes in GWAS studies.  These methods are implemented in our software package MPAT (Multiple Phenotype Association Tests).  Contrary to the common notion and practice of PCA analysis which usually retains the top few PCs that explain most of the variability in the data for dimension reduction to be used for testing for genetic effects with multiple phenotypes, we found that  the higher order PCs can be more powerful than the top PCs for association analysis. This counter-intuitive phenomenon can be well explained by the novel geometric concept of principal angle first introduced in this paper. Theoretically, a particular PC is powerful if its principal angle is zero and powerless if its principal angle is $90^o$. Prior to the introduction of the novel concept principal angle, the power of PC based tests for the multivariate normal means  depends on the mean vector ($K$ parameters) and the correlation matrix ($K(K-1)/2$ parameters). With the help of principal angle, the power of PC based tests only depends on the $K$ principal angles, $K$ eigenvalues and the overall signal strength. Hence, the complexity of power analysis for PC-based tests reduces from quadratic to linear order in the number of phenotypes. However, the principal angles are generally unknown in practical settings. One cannot choose a particular PC based on estimated principal angles and then use that cherry-picked PC for inference, because this approach will incur data snooping bias and the type I error rate will be inflated. Actually, the proposed PCMinP test correctly adjusts for this cherry-picking process and provides a valid inference. 

Effective combination  of PCs for multiple phenotype genetic association testing depends on the $K$ eigenvalues and  the $K$ unknown principal angles. We proposed linear, nonlinear and adaptive omnibus combinations of PCs to achieve robust power. 
PCLC is an inverse-eigenvalue weighted linear combination of PCs and can be as powerful as the Oracle test when all the principal angles are equal to each other, but can lose power otherwise. In the worst case, PCLC is powerless when the signal vector is parallel to its rejection boundaries. PCMinP is expected to perform well when there exists one principal angle equal to zero, but can lose power when the signal vector lies in the middle of all the PC directions.  The PCFisher test  combines all the mutually independent principal component $p$-values using Fisher's method, which can be more powerful than PCMinP when the signal vector lies in the middle of all the PCs but can be less powerful than PCMinP when some principal angle is equal to zero. We further proposed three quadratic combinations of PCs: WI, Wald and VC. Surprisingly, the classical Wald test and the variance component score test using the linear mixed model framework \citep{Huang2013,liu2017} are two special cases of weighted quadratic combinations of PCs.  Using convex optimization, we found that the Wald test achieves its maximal power when the last principal angle is zero and minimal power when the first principal angle is zero. The VC test is more powerful than the Wald test when the last principal angle is zero and can be less powerful otherwise. The WI test is more powerful than both the Wald and VC tests when the first principal angle is zero but can be less powerful otherwise. None of them is robust to the unknown principal angles.

The adaptive  quadratic test PCAQ is more robust than the WI, Wald and VC tests. As demonstrated by the simulation studies, PCMinP and PCLC can be more powerful than PCAQ in some situations. This  suggests that an omnibus test that combines all these six tests together would be even more robust than PCAQ.  The $p$-values of PCAQ and PCO can both be calculated analytically by numerical integration. This is advantageous when analyzing a large number of phenotypes with millions of SNPs across the whole genome,  as the principle angles are likely to change from one SNP to another and a powerful test for one SNP might not be powerful for another SNP.  All the proposed testing procedures  have been implemented in a publicly  available R package \textit{MPAT}. The connections and subtle differences between those PC based tests were illustrated graphically in terms of their rejection boundaries.   The theoretical conditions under which each PC based test can be more powerful than the traditional Wald test  are as follows: the principal angles $\theta_k= 0$ for  PC$_k$, PCMinP and PCO; $\cos^2(\theta_k)=1/K$ for PCLC; $\theta_1=0$ for WI; $\theta_K=0$ for VC;  $\theta_1=0$ or $\theta_K=0$   for PCAQ.

The eigen-analysis section investigates how the correlation structures  among multiple phenotypes can influence the eigenvalues and eigenvectors of the correlation matrix, and subsequently affect the powers of PC based tests.   From eigen-analysis and simulation studies, we found that the PCO test can outperform the MinP and the TATES tests for the detection of sparse signals, especially when the dimension is high.  The classical Wald test can perform poorly in high dimension settings as discovered by our eigen-analysis and demonstrated empirically by simulation studies, whereas the omnibus test PCO can still have good power in those settings. The eigen-analysis highlights the importance of the correlation structures in affecting the powers of PC based tests for detecting both sparse and dense signals. The eigen-analysis also shows that caution is needed for PC-based multiple phenotype analysis in the presence of highly correlated phenotypes. In such cases, the covariance matrix of multiple phenotypes is close to be singular, and the eigenvalues of the last few PCs are likely to be very small, making some tests that combine PCs, such as the Wald test, unstable. One can either remove some of highly correlated phenotypes before performing multiple phenotype tests, or remove the last few PCs with very small eigenvalues before combining PCs to construct tests. For the former, one can select biologically meaningful phenotypes for a joint analysis in collaboration with domain scientists. At the same time, statistical consideration of power and numerical stability also should be taken into account when selecting phenotypes into analysis.   Further research is needed on  how to  truncate PCs using selective inference theory  \citep{choi2014selecting} and then use those truncated PCs to construct valid and powerful tests by balancing the power and the numerical stability of the tests. 

In this post-GWAS era, there are increasing amounts of GWAS summary statistics for multiple phenotypes publicly available on dbGAP ({\em{http://www.ncbi.nlm.nih.gov/gap}}) and other places.  Our methods and software provide a cost-effective way  to analyze such data sets to discover novel biology by borrowing information across multiple phenotypes. We demonstrated the usefulness of our methods by analyzing multiple metabolic syndrome related clinical phenotypes with data sets collected from four international consortia. This real data example illustrates that the PCO test has robust power to detect additional novel loci underlying metabolic syndrome, outperforming the existing TATES method.  It is of future research interest to apply our tests to higher dimensional practical settings, for instance, in the studies of the genetic basis of gene expression levels  or DNA methylation levels in a biological pathway/network when such data sets are available. When individual level data are available for both phenotypes and genotypes in the future, it would be practically interesting to compare the performances of our PC based tests with other multiple phenotypes methods as discussed in \citep{Galesloot2014}.

PCA is just one dimension reduction method  for transforming the correlated Z-statistics into uncorrelated ones using spectral-decomposition of the correlation matrix. There  exist other methods for de-correlating correlated Z-statistics, such as the Cholesky decomposition. It would be interesting to explore the differences and connections between the testing statistics obtained from eigen-decomposition and Cholesky decomposition. With the increased availability of phenome-wide data,  there will be a greater demand for analyzing multiple phenotypes in sequencing studies especially using electronic medical record data and molecular phenotype data. Since popular region-based  association testing statistics for rare variants  are not normally distributed, for example, the SKAT test statistic follows a mixture of chi-squared distributions \citep{Lee2012}, our current PC based tests are not directly applicable for the detection of associations between rare variants and  multiple phenotypes.  More future work is needed to extend  the current PCA framework for joint analysis of multiple phenotypes in GWAS to multiple phenotype analysis in sequencing association studies to detect rare variant effects.

\vspace{-0.3in}
\begin{singlespace}
\bibliographystyle{biom}
\bibliography{PCA}

\end{singlespace}

\section*{Supplementary Materials}
The supplementary pdf file contains simulation results and additional real data analysis results. The supplementary excel file contains the unstructured correlation matrix of dimension $100\times 100$, and the $\bbeta$ vectors of length $100$ in simulation settings M11-M15, and the estimated principal angles for the ten SNPs in Table \ref{Table_MetS-pvalues}.

\end{document}


\title{{\bf Supplemental Material for \\
``A Geometric Perspective on the Power of Principal Component Association Tests in Multiple Phenotype Studies''}
}
\author{Zhonghua Liu and Xihong Lin}
\date{}
\maketitle

\section*{S1.Simulation Studies}
\begin{figure}[!p]
\centering
      \includegraphics[scale=0.7]{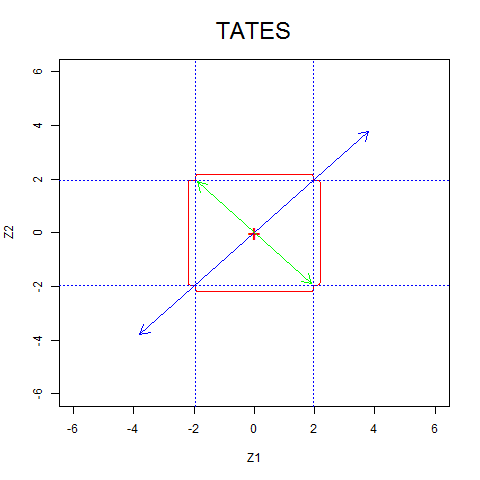}
        \caption{The rejection boundary of the TATES method}\label{fig:fig_s1}
\end{figure}

The rejection boundary of the TATES method is givne in Figure \ref{fig:fig_s1}.

\section*{S2 Joint Analysis of the MetS Related Phenotypes}
In this section, we first provide the web links for the data sets used in the main text. The GIANT consortium data sets can be downloaded at \url{http://portals.broadinstitute.org/collaboration/giant/index.php/GIANT_consortium_data_files} in the sections of GWAS Anthropometric 2015 BMI and GWAS Anthropometric 2015 Waist. The lipids data sets can be downloaded at \url{http://csg.sph.umich.edu/abecasis/public/lipids2010/} and \url{http://csg.sph.umich.edu/abecasis/public/lipids2013}. The MAGIC data sets are at \url{https://www.magicinvestigators.org/downloads/} in the section of Glucose and insulin results accounting for BMI. The ICBP data sets are  located at \url{https://www.ncbi.nlm.nih.gov/projects/gap/cgi-bin/study.cgi?study_id=phs000585.v1.p1}. 

We first merged those raw data sets by the chromosome:position column and then apply the proposed PC methods and the TATES method to the Z-scores of each phenotype across all the SNPs.  We further calculated genomic control factors for the p-values of the original univariate analysis performed by the four consortia, PC based tests and the TATES methods, and we didn't observe any serious inflation with genomic control factor in the range of $(0.92,1.13)$. The QQ plots for those p-values are also provided in Figure S1-S8. Due to numerical precision of R software, the 
p-values of PCMinP, PCAQ and PCO are truncated at zeros if smaller than certain thresholds. 

\begin{figure}[!p]
\centering
    \begin{subfigure}{.45\textwidth}
        \centering
        \includegraphics[width=\linewidth]{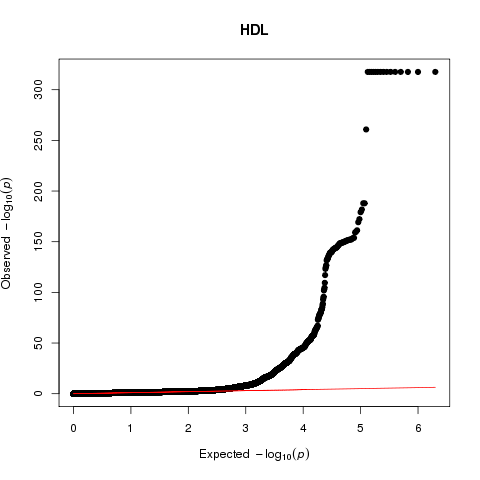}
        \caption{HDL}\label{fig:fig_a}
    \end{subfigure} %
    \begin{subfigure}{.45\textwidth}
        \centering
        \includegraphics[width=\linewidth]{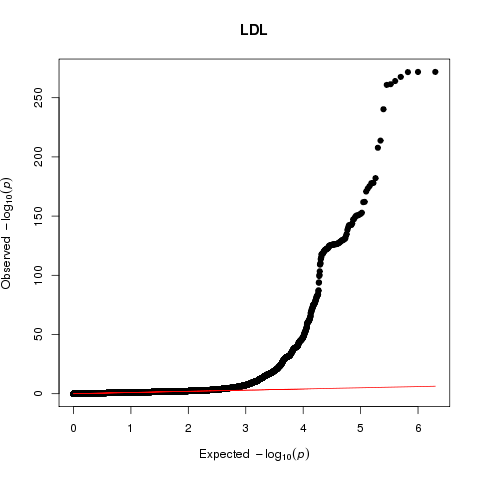}
        \caption{LDL}\label{fig:fig_b}
    \end{subfigure} %
    \begin{subfigure}{0.45\textwidth}
        \centering
        \includegraphics[width=\linewidth]{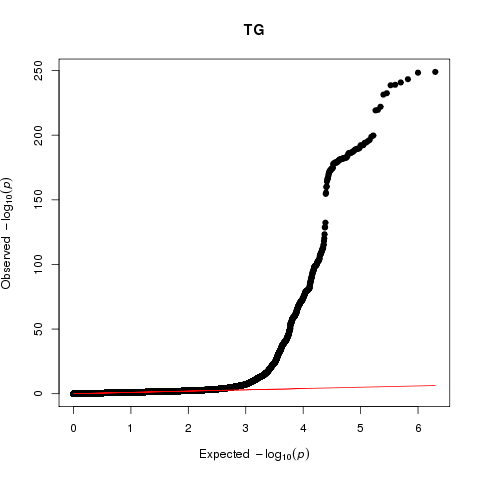}
        \caption{TC}\label{fig:fig_c}
    \end{subfigure}
      \begin{subfigure}{0.45\textwidth}
        \centering
        \includegraphics[width=\linewidth]{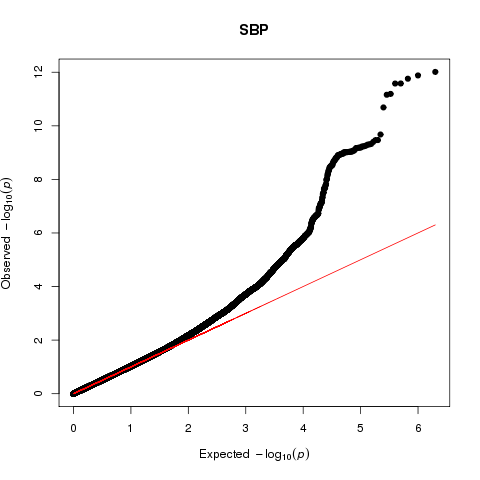}
        \caption{TG}\label{fig:fig_d}
    \end{subfigure}
\caption{QQ plots for the p-values of the original univariate analysis of HDL, LDL and TG by the GLGC, and SBP by ICBP}
\label{fig:s1}
\end{figure}

\begin{figure}[!p]
\centering
    \begin{subfigure}{.45\textwidth}
        \centering
        \includegraphics[width=\linewidth]{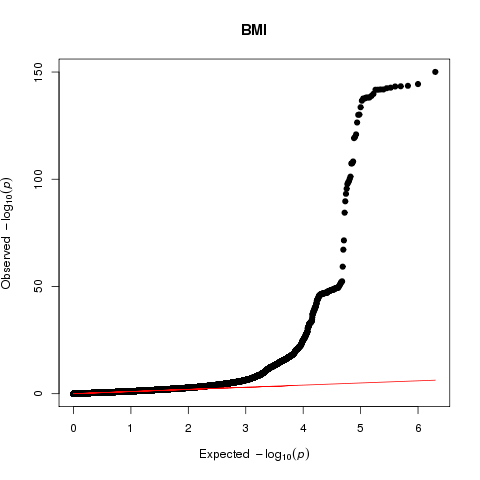}
        \caption{BMI}\label{fig:fig_a2}
    \end{subfigure} %
    \begin{subfigure}{.45\textwidth}
        \centering
        \includegraphics[width=\linewidth]{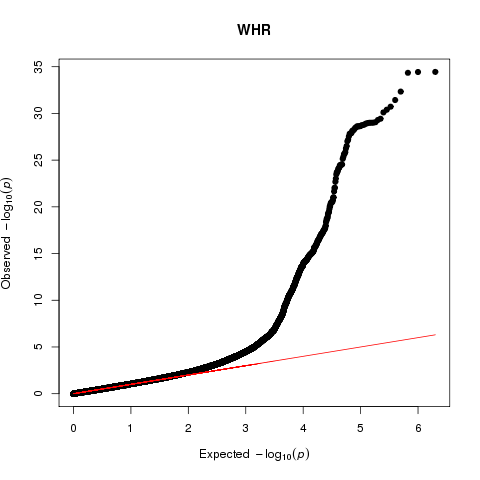}
        \caption{WHR}\label{fig:fig_b2}
    \end{subfigure} %
\caption{QQ plots for the p-values of the original univariate analysis of BMI and WHR by the GIANT.  }
\label{fig:s3}
\end{figure}

\begin{figure}[!p]
\centering
    \begin{subfigure}{.45\textwidth}
        \centering
        \includegraphics[width=\linewidth]{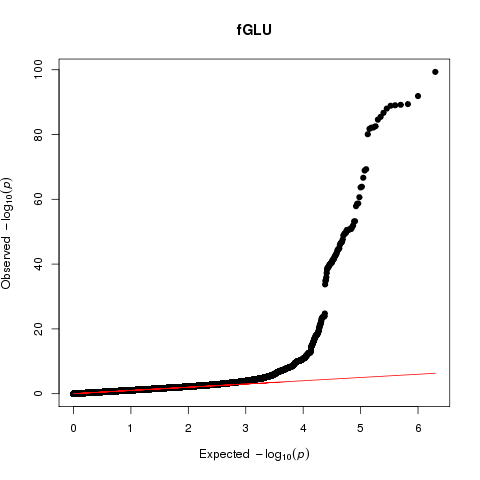}
        \caption{FG}\label{fig:fig_a3}
    \end{subfigure} %
    \begin{subfigure}{.45\textwidth}
        \centering
        \includegraphics[width=\linewidth]{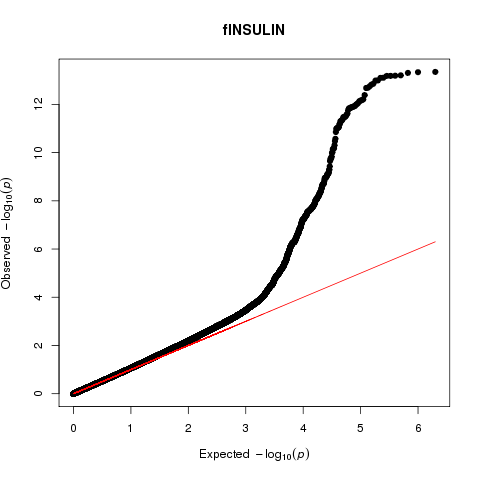}
        \caption{FI}\label{fig:fig_b3}
    \end{subfigure} %
\caption{QQ plots for the p-values of the original univariate analysis of FG and FI by the MAGIC.  }
\label{fig:s4}
\end{figure}

\begin{figure}[!p]
\centering
    \begin{subfigure}{.45\textwidth}
        \centering
        \includegraphics[width=\linewidth]{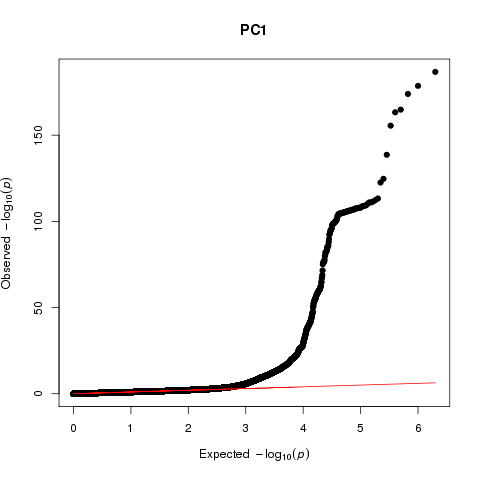}
        \caption{PC1}\label{fig:fig_a4}
    \end{subfigure} %
    \begin{subfigure}{.45\textwidth}
        \centering
        \includegraphics[width=\linewidth]{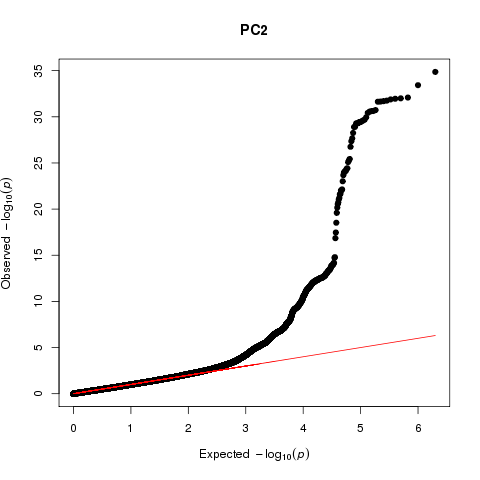}
        \caption{PC2}\label{fig:fig_b4}
    \end{subfigure} %
    \begin{subfigure}{0.45\textwidth}
        \centering
        \includegraphics[width=\linewidth]{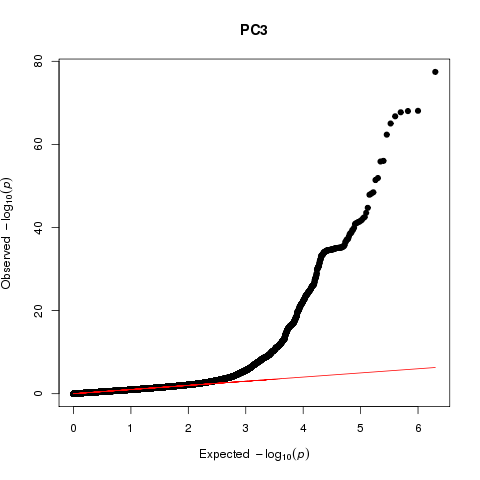}
        \caption{PC3}\label{fig:fig_c4}
    \end{subfigure}
      \begin{subfigure}{0.45\textwidth}
        \centering
        \includegraphics[width=\linewidth]{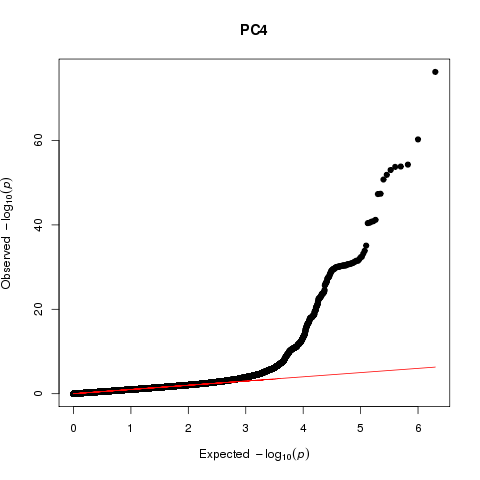}
        \caption{PC4}\label{fig:fig_d4}
    \end{subfigure}
\caption{QQ plots for the p-values of PC1, PC2, PC3 and PC4.}
\label{fig:s5}
\end{figure}

\begin{figure}[!p]
\centering
    \begin{subfigure}{.45\textwidth}
        \centering
        \includegraphics[width=\linewidth]{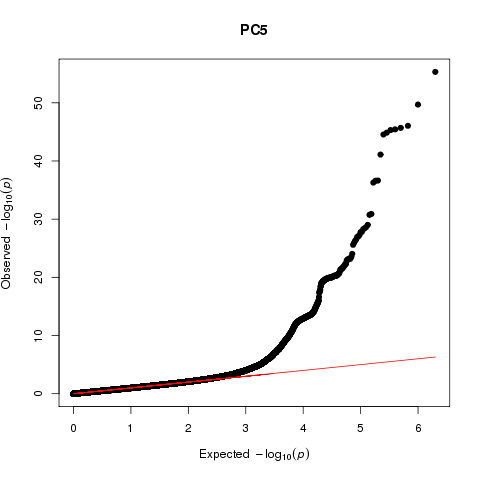}
        \caption{PC5}\label{fig:fig_a5}
    \end{subfigure} %
    \begin{subfigure}{.45\textwidth}
        \centering
        \includegraphics[width=\linewidth]{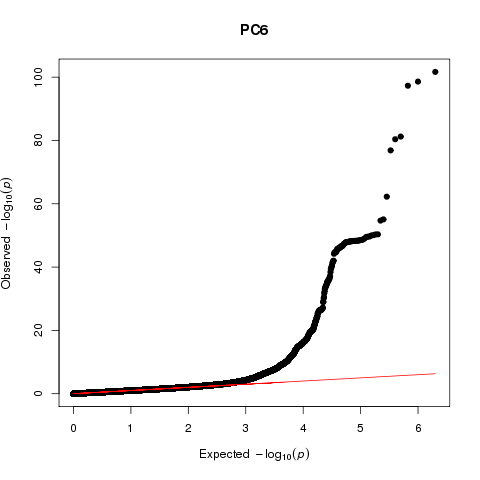}
        \caption{PC6}\label{fig:fig_b5}
    \end{subfigure} %
    \begin{subfigure}{0.45\textwidth}
        \centering
        \includegraphics[width=\linewidth]{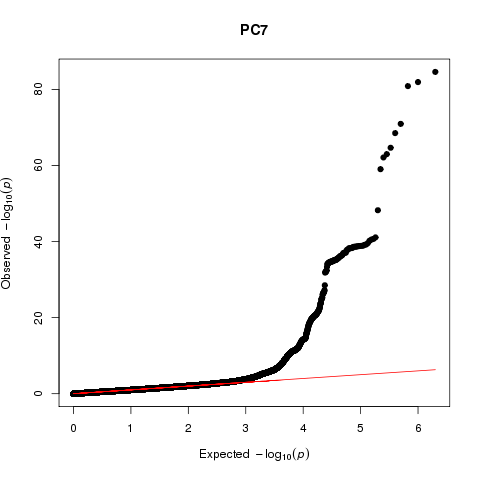}
        \caption{PC7}\label{fig:fig_c5}
    \end{subfigure}
      \begin{subfigure}{0.45\textwidth}
        \centering
        \includegraphics[width=\linewidth]{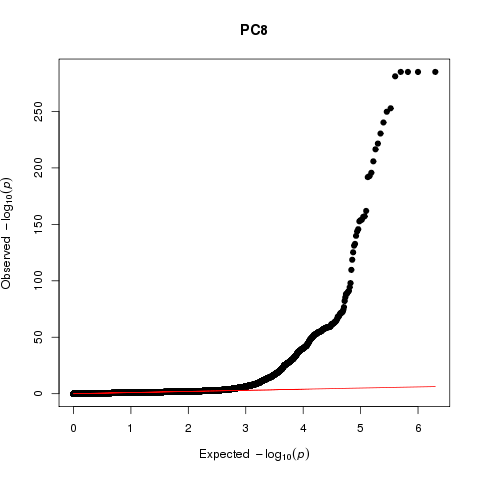}
        \caption{PC8}\label{fig:fig_d5}
    \end{subfigure}
\caption{QQ plots for the p-values of PC5, PC6, PC7, PC8.}
\label{fig:s6}
\end{figure}

\begin{figure}[!p]
\centering
        \begin{subfigure}{0.45\textwidth}
        \includegraphics[width=\linewidth]{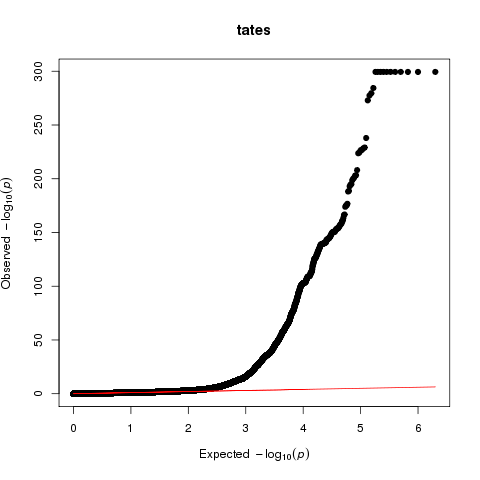}
        \caption{TATES}\label{fig:fig_a6}
    \end{subfigure}
       \begin{subfigure}{.45\textwidth}
        \centering
        \includegraphics[width=\linewidth]{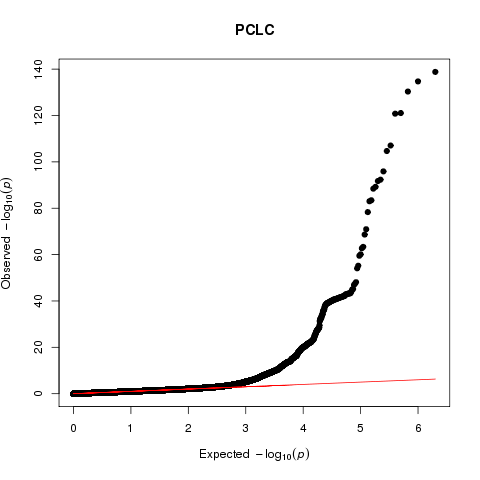}
        \caption{PCLC}\label{fig:fig_c6}
    \end{subfigure} %
    \begin{subfigure}{.45\textwidth}
        \centering
        \includegraphics[width=\linewidth]{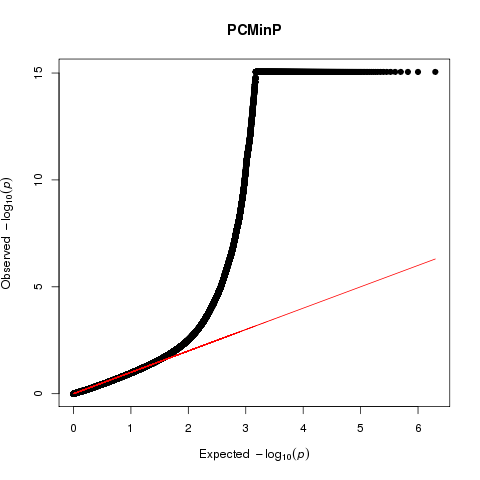}
        \caption{PCMinP}\label{fig:fig_d6}
    \end{subfigure} %
    \begin{subfigure}{0.45\textwidth}
        \centering
        \includegraphics[width=\linewidth]{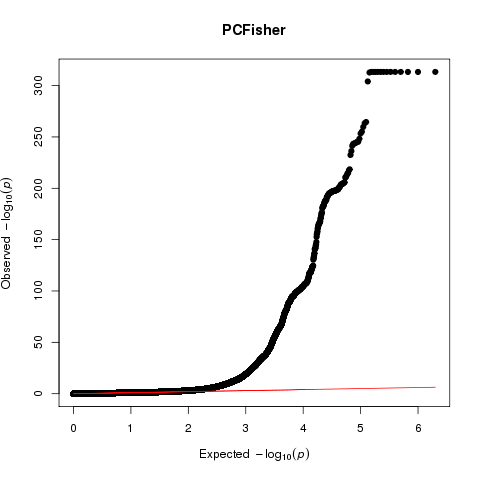}
        \caption{PCFisher}\label{fig:fig_e6}
    \end{subfigure}
      \begin{subfigure}{0.4\textwidth}
        \centering
        \includegraphics[width=\linewidth]{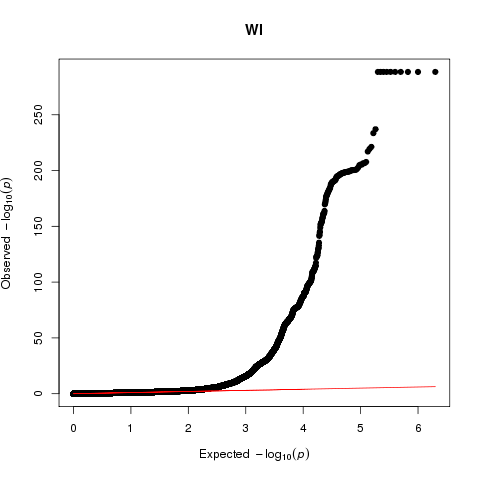}
        \caption{WI}\label{fig:fig_f6}
    \end{subfigure}
\caption{QQ plots for PC9, TATES, PCLC, PCMinP, PCFisher and WI.}
\label{fig:s7}
\end{figure}

\begin{figure}[!p]
\centering
    \begin{subfigure}{.45\textwidth}
        \centering
        \includegraphics[width=\linewidth]{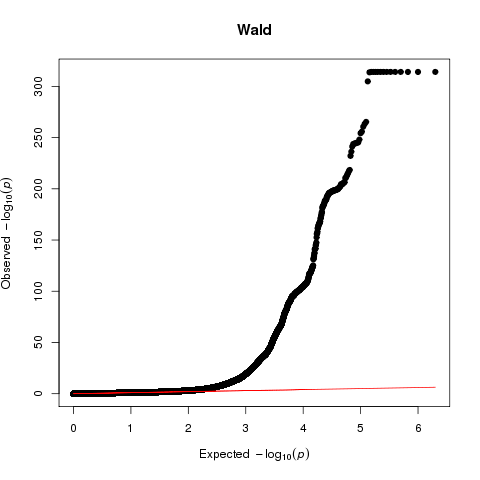}
        \caption{Wald}\label{fig:fig_a7}
    \end{subfigure} %
    \begin{subfigure}{.45\textwidth}
        \centering
        \includegraphics[width=\linewidth]{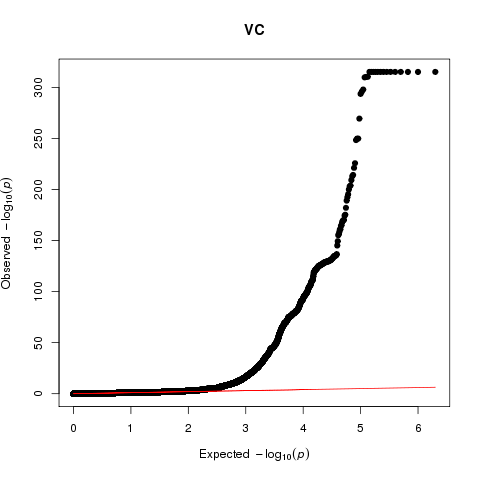}
        \caption{VC}\label{fig:fig_b7}
    \end{subfigure} %
    \begin{subfigure}{0.45\textwidth}
        \centering
        \includegraphics[width=\linewidth]{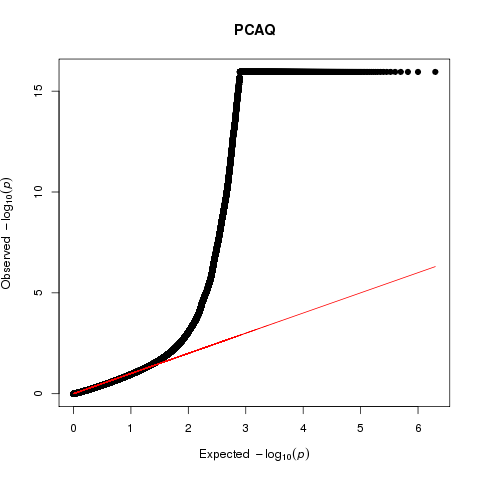}
        \caption{PCAQ}\label{fig:fig_c7}
    \end{subfigure}
      \begin{subfigure}{0.45\textwidth}
        \centering
        \includegraphics[width=\linewidth]{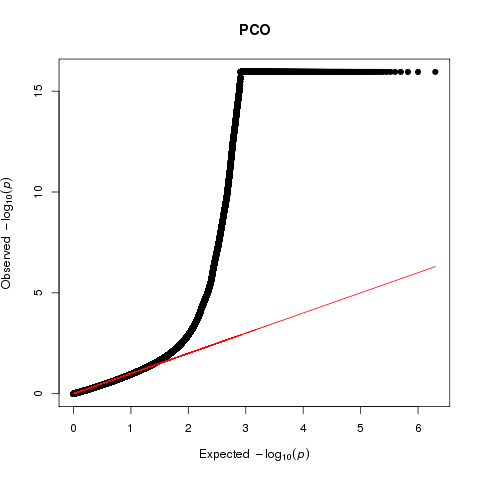}
        \caption{PCO}\label{fig:fig_d8}
    \end{subfigure}
\caption{QQ plots for the p-values of Wald, VC, PCAQ, PCO.}
\label{fig:s8}
\end{figure}

